\documentclass[twocolumn,pre,showpacs,amsfonts,amsmath,assymb,eufrak
,longbibliography]{revtex4-2}
\usepackage{epsfig}
\usepackage{color}
\usepackage{bm}
\usepackage[
colorlinks=true, 
pdfstartview=FitV, 
linkcolor=blue, 
citecolor=magenta, 
urlcolor=blue, 
bookmarks=true,
bookmarksnumbered=true,
pdftitle={},
pdfauthor={}
]{hyperref}
\usepackage{epsfig,amsopn}
\usepackage{graphicx}

\usepackage{amssymb}
\usepackage{amsmath}
\usepackage{color}
\usepackage{subfigure}
\usepackage{textcomp}
\usepackage[utf8]{inputenc}
\definecolor{pink}{rgb}{1,0,0.6}

\newcommand\nn{\nonumber}

\usepackage{grffile}
\graphicspath{{./Figures/}{./Hydro/}{./SpaceAverage/}}
\usepackage{textcomp}

\newcommand{\be}{\begin{equation}}
\newcommand{\ee}{\end{equation}}
\newcommand{\bea}{\begin{eqnarray}}
\newcommand{\eea}{\end{eqnarray}}

\begin{document}

\newcommand{\titlename}{Density-dependent transport coefficients in two-dimensional cellular aggregates}

\title{\titlename}

\author{Subhadip Chakraborti$^{1,2}$ and Vasily Zaburdaev$^{1,2}$}

\affiliation{$^1$Department of Biology, Friedrich-Alexander-Universit\"{a}t Erlangen-N\"{u}rnberg, 91058 Erlangen, Germany}%
\affiliation{$^2$Max Planck Zentrum für Physik und Medizin, 91054 Erlangen, Germany}

\date{\today}
\begin{abstract} 
The large-scale collective behavior of biological systems can be characterized by macroscopic transport, which arises from the non-equilibrium microscopic interactions among individual constituents. A prominent example is the formation of dynamic aggregates by motile eukaryotic cells or bacteria mediated by active contractile forces. In this work, we develop the two-dimensional fluctuating hydrodynamics theory based on the microscopic dynamics of a model system of aggregation by \textit{Neisseria gonorrhoeae} bacteria. The derivation of two macroscopic transport coefficients of bulk diffusivity and conductivity which determine hydrodynamic current of cells is the central result of this work. By showing how transport coefficients depend on cell density and microscopic parameters of the system we  predict transport slowdown during the colony formation process. This study provides valuable analytical tools for quantifying hydrodynamic transport in experimental systems involving cellular aggregation occurring due to intermittent contractile dipole forces.
\end{abstract}
\maketitle

\section{Introduction}
\label{intro}

Active motility is common to individual eukaryotic and prokaryotic cells and driven by forces exerted by cells on their microenvironment -- either a substrate or an extracellular matrix \cite{Gutnick1998, Zaburdaev2015, Wolfgang_Review1987, Friedl_NatureReview2003, Fabry_PLoS2012, Fabry_eLife2020, Karow_Nature2018, Karow_2021, Lancaster_Nature2013, Pasca_Nature2018}. Furthermore, if multiple motile cells are able to  pull on each other forming a dipole-like attractive interaction, they can form multicellular aggregates \cite{Zaburdaev2017, Fernandez2021, Pasca2024, Wang2025}. These interactions are often short range and are sustained up to a random persistence time. This active motility and interactions drive the system far from thermal equilibrium and thus may produce rich non-equilibrium phenomena such as formation and merging of micro-colonies, ultimately leading to dense cellular aggregates \cite{Sanchez2012, Alert2020}.

One important quantity to characterize the non-equilibrium nature of any system is the current \cite{Onsager, Pathria, Bertini_RMP2015}. The dynamical state before reaching the steady state is always out of equilibrium and hence shows a non-zero current. Whereas the time stationary state can either be equilibrium or non-equilibrium depending on the presence of the current. Depending on the system, this current could be heat current, charge current or particle current. Sometimes the current is not readily visible in the space of chosen coordinates in non-equilibrium steady state (NESS) but it is always present in the phase space manifesting breaking of the \textit{detailed balance} \cite{Gardinerbook, Broedersz2018, Broedersz2024}. 

Interestingly, this current not only indicates the out of equilibrium nature of the system but also provides important information on the transports (such as diffusion and conductivity) within the system. For biological systems, the transport has practical association with the survival of the colony through controlling the access to nutrients, waste disposal, and collective response to external perturbations \cite{Ascione2024, Xu2019}. Thus the macroscopic biological function of cellular aggregates such as phenotype differentiation in bacterial colonies \cite{ponisch2018pili}, cell fate specification in organoids \cite{tortorella2021role} or developing antibiotic resistance by controlling transport within bacterial colonies \cite{Maier2021}, is largely influenced by the microscopic active fluctuations contributed by individual cells. However, a systematic and detailed formulation of macroscopic transport arising from the microscopic active fluctuation is less known and emerged very recently through the derivation of fluctuating hydrodynamics. The derivation of fluctuating hydrodynamics, on one hand, describes large scale spatio-temporal behavior of the system by characterising hydrodynamic currents, on the other hand, provides detailed information on the dependence of the coarse-grained transport coefficients on the key active microscopic parameters. 

In this paper, we develop a theoretical framework to derive macroscopic transport coefficients by constructing a fluctuating hydrodynamics description for a broad class of cellular aggregates interacting via intermittent active dipolar forces with short (exponential) interaction ranges. While our study is broadly applicable, our model system is motivated by features observed in \textit{Neisseria gonorrhoeae} (\textit{N. gonorrhoeae}) bacteria, which utilize dynamic, pili-mediated forces to form dense clusters. To capture the essential aspects of such interactions, we introduce a two-dimensional lattice-based model incorporating key microscopic features of short-range contractile forces \cite{Tainer2004, Mattick2002, Maier_PRL2010, SheetzMP_Nature2000, Zaburdaev2017, Berg_PNAS2001, Maier2019}. Through this model, we observe a clustering transition controlled by the balance between cell-cell and cell-substrate interactions. 
We then derive fluctuating hydrodynamics for the particle (cell) density using a recently developed macroscopic fluctuation theory \cite{Bertini_PRL2001, Bertini_RMP2015, Derrida_JSM2007}. This formalism leads to a fluctuating coarse-grained current which involves two density dependent transport coefficients namely bulk-diffusivity and conductivity. This work builds on our earlier semi-analytical study \cite{Chakraborti-PRR2024} of the one-dimensional system by extending the framework to two dimensions and providing complete analytical expressions for the transport coefficients. Our analytical expressions for this transport coefficients can capture the results calculated from 2D agent-based simulations quite well even in the highly non-equilibrium regime where the clustering is very prominent due to high cell-cell interaction rate. 
We argue that both bulk-diffusivity and conductivity can be measured from experiments with cell aggregates as proportionality factors of the hydrodynamic cell-current through an imaginary coarse grained volume within the colony in the presence of a density gradient or an external force, respectively. 

We would like to emphasize that a central challenge in the theoretical study of active matter is the derivation of analytical expressions for transport coefficients, such as the density-dependent diffusion coefficient, in interacting systems. Due to their intrinsically far-from-equilibrium nature, active matter models with interactions rarely admit closed-form analytical solutions. In many cases, effective hydrodynamic theories incorporate phenomenological density-dependent diffusion terms to describe clustering and motility-induced phase separation (MIPS) \cite{Cates_review}. Here, we develop a fluctuating hydrodynamics theory that explicitly derives an analytical expression for diffusivity and conductivity. This analytical progress offers a solid quantitative insight into the microscopic origins of density-dependent transport in active systems with complex interactions.

While the present study builds upon the framework developed in Ref.~\cite{Chakraborti-PRR2024}, it introduces several important advancements that significantly extend its scope and applicability. Most notably, the model is extended from one to two (and potentially higher) dimensions thus opening up its direct comparison to the real experimental settings. Taking advantage of closed form expression for particle-particle interactions we were also able to find an analytical description of probability density distributions required for the derivation of the hydrodynamic equations, and thus enabling a complete and explicit determination of the transport coefficients -- which was not achieved in the previous study.

The paper is organized as follows. In Sec.~\ref{sec:model}, we define the  stochastic model of aggregation on two-dimensional lattice and characterise the clustering transition. In Sec.~\ref{sec:hydro}, we map the exclusion model (which our lattice model effectively is) to an unbounded model and derive a fluctuating hydrodynamic description starting from microscopic dynamics and obtain the transport coefficients analytically. Sec.~\ref{sec:result} shows the numerical results and comparison with the theory. We conclude in Sec.~\ref{conclusion} with the summary of the main results and a discussion of the remaining open questions.

\begin{figure}
\begin{center}
\includegraphics[width=7.8cm,angle=0]{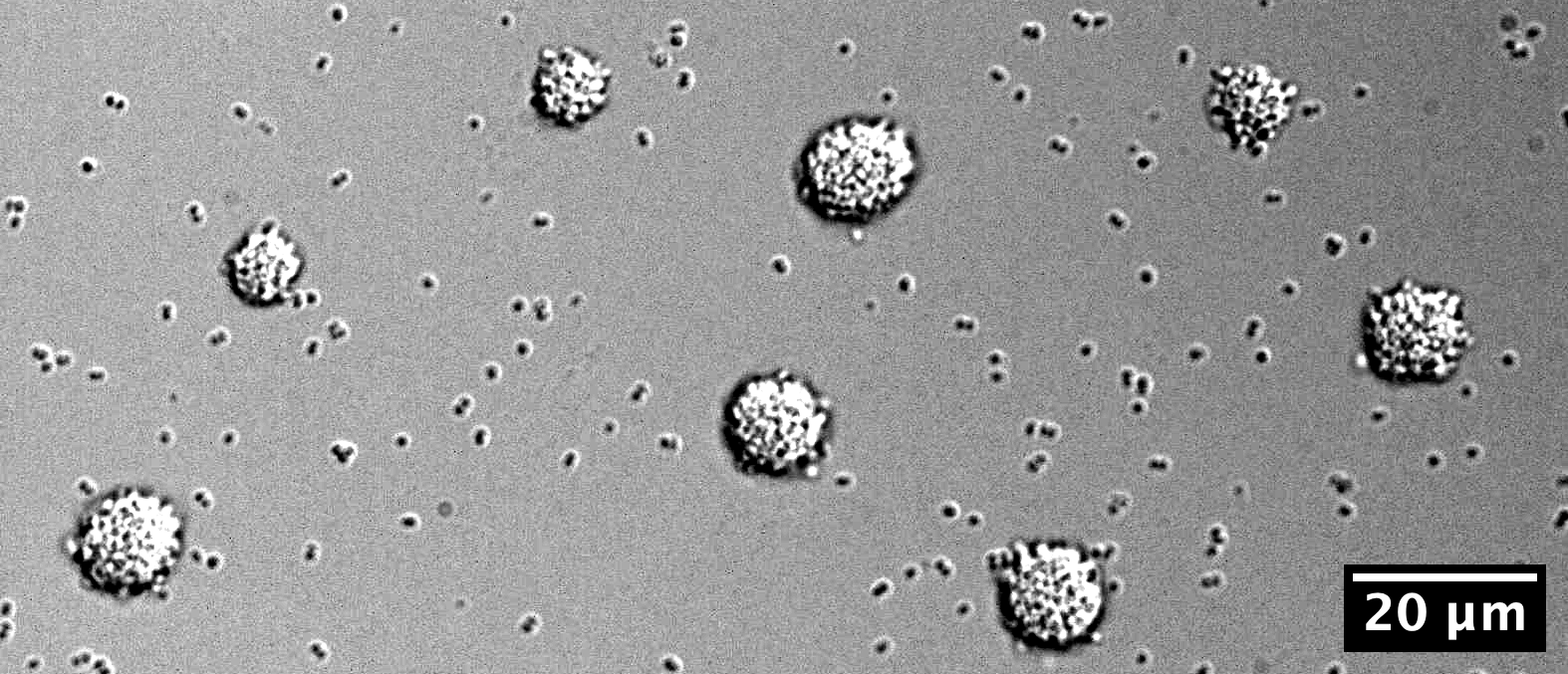}
\put (-220,85) {{(a)}}

\vspace{0.1cm}

\includegraphics[width=8cm,angle=0]{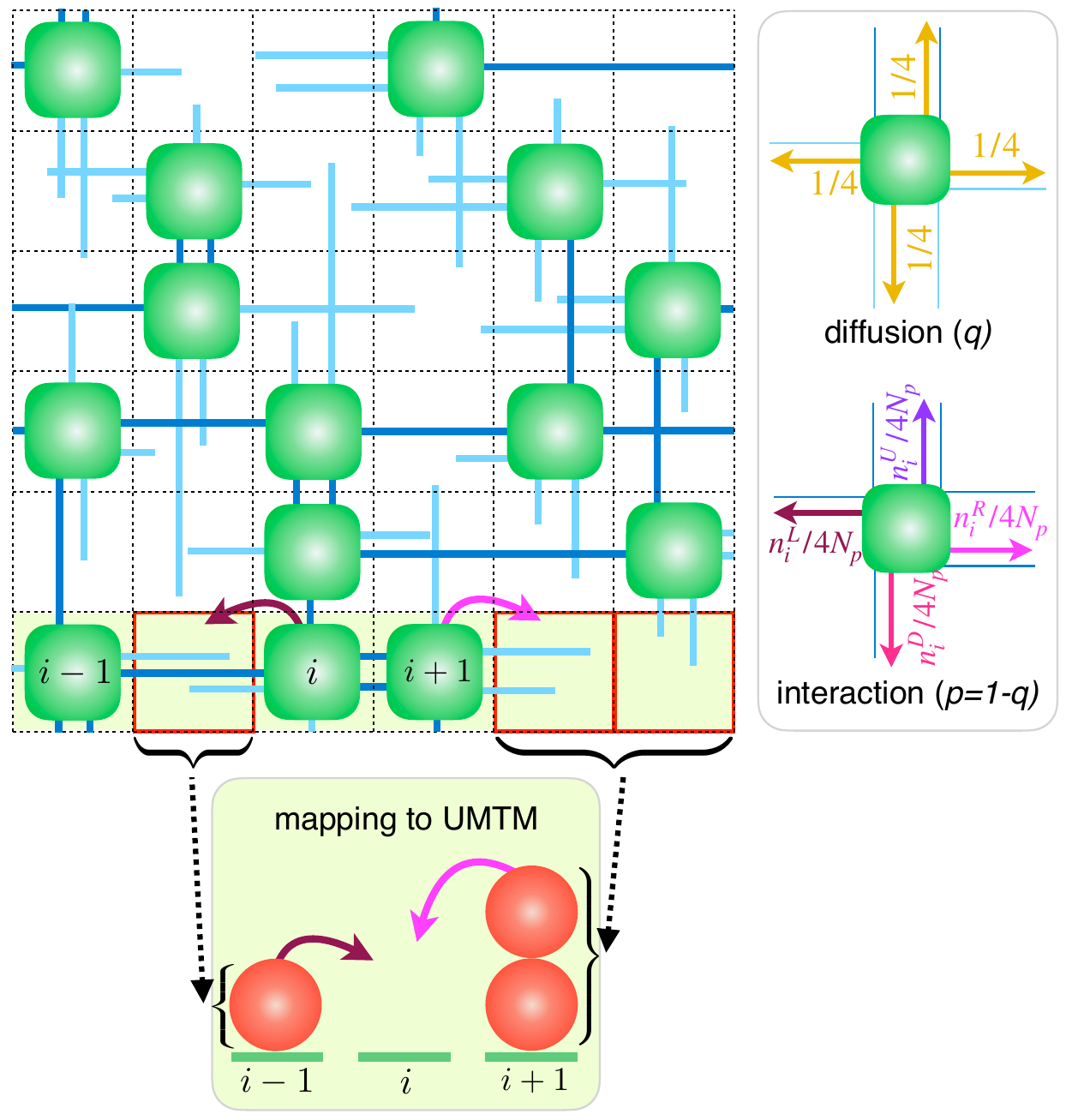}
\put (-225,240) {(b)}
\put (-64,240) {(c)}
\put (-200,40) {(d)}

\caption{Schematic diagram for the microscopic interactions of \textit{N. gonorrhoeae} bacteria. Panel (a): Differential interference contrast (DIC) image of a formation of \textit{N. gonorrhoeae} bacterial aggregates on a solid substrate (see \cite{Zaburdaev2015} for experimental details and video of the aggregation process). Large aggregates are surrounded by isolated individual cells. Image courtesy of N. Biais. Panel (b): In the two-dimensional exclusion model (EM) with periodic boundary, the free pili are represented by light blue lines and the bounded pairs are shown by dark blue one. Panel (c): Uniform hopping rates due to pili-substrate interaction (occurring with stochastic weight $q$) are indicated by yellow arrows. Hopping due to the bound-pair dependent cell-cell interactions (occurring with stochastic weight $p=1-q$) are represented by four different colors. Panel (d) shows the mapping of the last row of the 2D EM to an unbounded mass transfer model (UMTM); other rows and columns can be mapped similarly. In this mapping, the adjacent vacancies in EM become masses in UMTM with a reverse direction of movements.}
\label{Schem1}
\end{center}
\end{figure}

\section{Model}
\label{sec:model}
We consider a model inspired by colony-forming \textit{N. gonorrhoeae} bacteria, as a representative example of a broader class of cellular aggregates that interact via short-ranged, intermittent active dipolar forces.

Many bacterial species, including \textit{N. gonorrhoeae} utilize retractable thin filaments known as type IV pili for environmental interactions and communication with each other \cite{Tainer2004, Mattick2002}.
Each bacterium has approximately 10-20 pili evenly distributed across its surface \cite{Zaburdaev2017, Maier_PRL2010}. Polymerisation (de-polymerisation) of pilin protein subunits leads to an extension (retraction) of a pilus filament with an exponential distribution of lengths, averaging around $1-2~\mu$m, with the cell size of about $1~\mu$m \cite{SheetzMP_Nature2000,Zaburdaev2017,Berg_PNAS2001,Maier2019}. When pili attach to a substrate and retract, they generate the pulling force necessary for twitching cell motility \cite{Mattick2002,Ellison,Henrichsen}. Stochastic binding and retraction of pili between neighboring cells generate an attractive force dipole, facilitating the formation of cell clusters [see Fig.~\ref{Schem1}(a)]. The characteristic scale of the pili-generated force is $\sim 180~$pN. The random detachment of pili affects the force balance by changing both the number of pili pairs bound to each other and the number of pili attached to the substrate and thus drives the movement of the cell.

We now construct a two-dimensional model to replicate the dynamics described above. Individual bacteria are represented by $N$ hardcore particles that can move on a 2D periodic lattice of size $L \times L$ [see Fig.~\ref{Schem1}(b)]. Since each lattice point can be occupied by at most one particle, these types of models are referred to as \textit{exclusion models} (EM) \cite{Evans_JPA2005}. Each particle is allocated with $N_p$ pili, randomly oriented in four directions: left, right, up, and down. The pili lengths follow an exponential distribution with a mean length $l_0$. Every pilus has an exponentially distributed lifetime with mean $T_0$. Upon the expiration of this lifetime, the pilus (whether free or bound) is destroyed and a new one is randomly assigned to one of the four directions on the same particle, maintaining the total number $N_p$ of pili per particle conserved. Pili dynamics affects both, particle motion on a substrate (with stochastic weight $q$) and particle-particle interactions leading to clustering (with weight $p=1-q$). A particle can hop one unit of lattice spacing $\ell$. 

We simplify the pili-substrate interaction to a symmetric hopping with probability $1/4$ [see Fig.~\ref{Schem1}(c)]. The particle dynamics resulting from cell-cell interactions are more complex. If the total length of two pili pointing towards each other from two adjacent particles is greater or equal to the distance between them then the pili will form a bound pair. A horizontal (vertical) pili can make pair only with a horizontal (vertical) pili; cross-linking between horizontal and vertical pili is not permitted. As shown in Fig.~\ref{Schem1}(c), a particle jumps toward left, right, up and down according to random weight $n^L/(4N_p)$,  $n^R/(4N_p)$, $n^U/(4N_p)$ and $n^D/(4N_p)$, respectively where $n^k$s are the number of bound pili pairs in the $k$-th direction.

This model system involves two key driving forces: pili-driven motility and intermittent cell-cell dipolar interactions. The pili-substrate interactions, without the simplification to symmetric hopping, result in effective persistent motion \cite{Zaburdaev_PRE2015, Zaburdaev_PRL2021}, leading to a qualitative alignment with an established active matter framework of motility-induced phase separation (MIPS) \cite{Cates_2013, Caprini_2020, Caporusso_2020}. However, unlike classical MIPS models that typically involve self-propelled particles with purely repulsive interactions, clustering in our system arises from intermittent, attractive, dipolar interactions mediated by pili.
The standard MIPS scenario -- arising purely from persistent motion, such as that generated by pili-substrate interactions -- is now relatively well understood \cite{Cates_2013, Caprini_2020, Caporusso_2020}, and has also been examined in the context of fluctuating hydrodynamics of run-and-tumbling particles \cite{Rahul_PRE2020}. Therefore, in this work, we simplify the cell-substrate interaction as symmetric diffusion and instead focus on the less explored role of active, pili-mediated cell-cell interactions in driving cluster formation and transport—distinguishing our model from previous active matter studies.

Interestingly, even when considering attraction alone, our model differs fundamentally from earlier studies of attractive active matter systems \cite{Redner_2013, Caprini_2023} which have considered models with passive attractive forces, such as Lennard-Jones potentials, where activity is introduced through an additional self-propulsion velocity. Rather, our model bears closer resemblance to the work of Ref.~\cite{Rafael_PRE2022} which investigated the role of intermittent dipolar attractions via an active switching process in microphase separation in a non-motile bacterial colony. It is important to note that the dipolar attraction in our model originates from intrinsically active interactions mediated by dynamic pili. As a result, the attraction itself is non-conservative and actively driven, giving rise to a form of non-equilibrium clustering that cannot be captured by conservative attractive potentials alone. The biologically motivated mechanism in our model leads to aggregation through a distinct route, although the resulting macroscopic behavior may resemble MIPS phenomenologically.

\begin{figure}
\begin{center}
\includegraphics[width=8cm,angle=0]{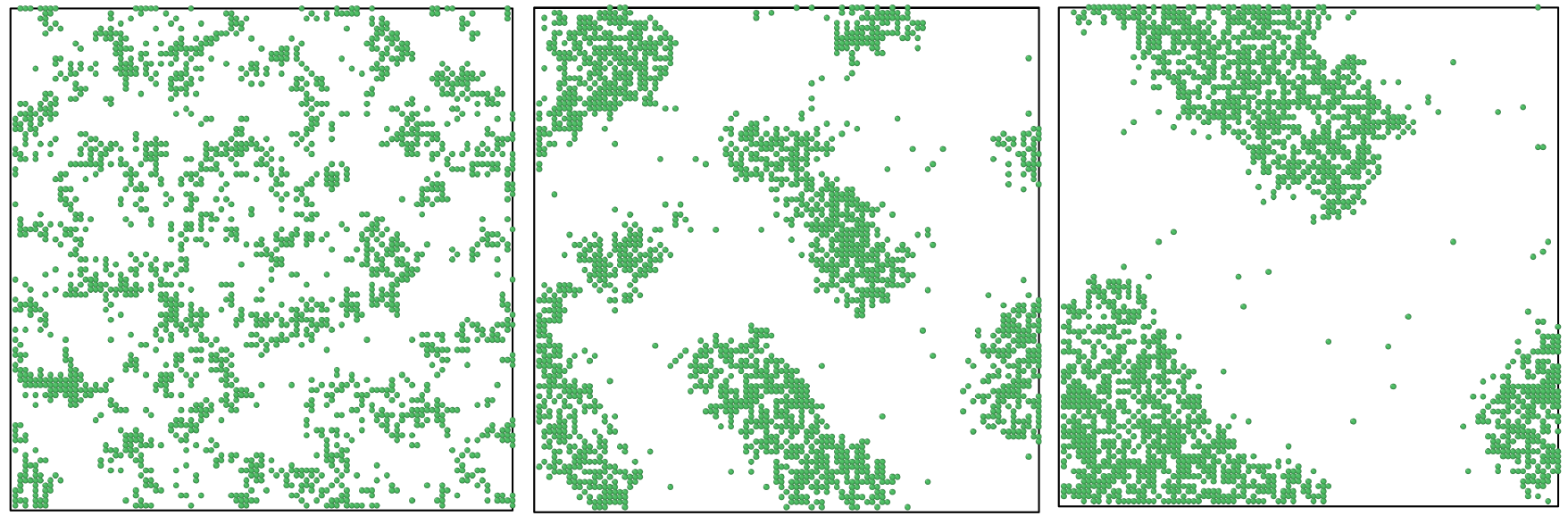}
\put (-227,77) {(a)}
\put (-215,-10) {$t=20\times L^2$}
\put (-132,-10) {$t=2\times L^3$}
\put (-50,-10) {$t=L^4$}

\vspace{0.1cm}
\includegraphics[width=8cm,angle=0]{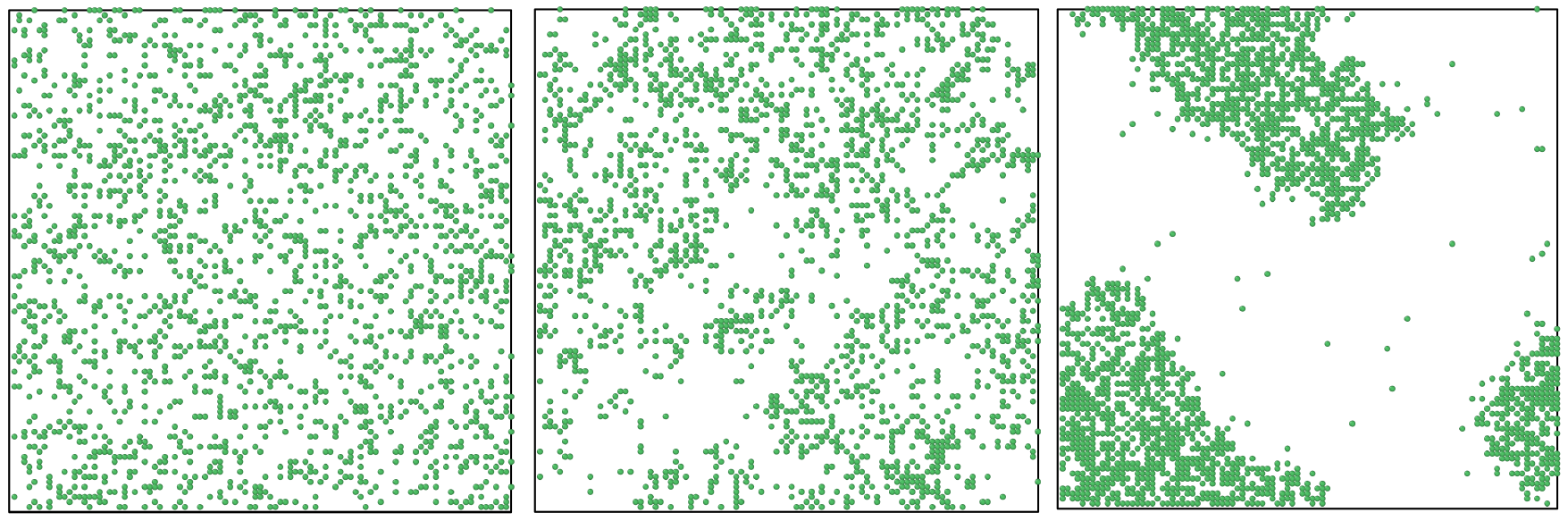}
\put (-227,77) {(b)}
\put (-205,-10) {$p=0.7$}
\put (-130,-10) {$p=0.9$}
\put (-55,-10) {$p=0.95$}
\caption{Clustering transition. We consider $N=2000$ particles in the 2D exclusion model of size $100 \ell \times 100 \ell$. Other parameters: Number of pili per particles $N_p=20$, mean lifetime of pili $T_0=10$ MCT and mean pili length $l_0=2 \ell$ with $\ell=1$ being the lattice spacing. Panel (a) shows, for the cell-cell interaction parameter $p=0.95$, the formation of microcolonies with time which eventually merge to form a big cluster surrounded by a low density gas of individual particles. In panel (b), we check the clustering at long time $t=L^4$ MCT for different values of $p$.}
\label{cluster}
\end{center}
\end{figure}

\begin{figure}
\begin{center}
\includegraphics[width=4.27cm,angle=0]{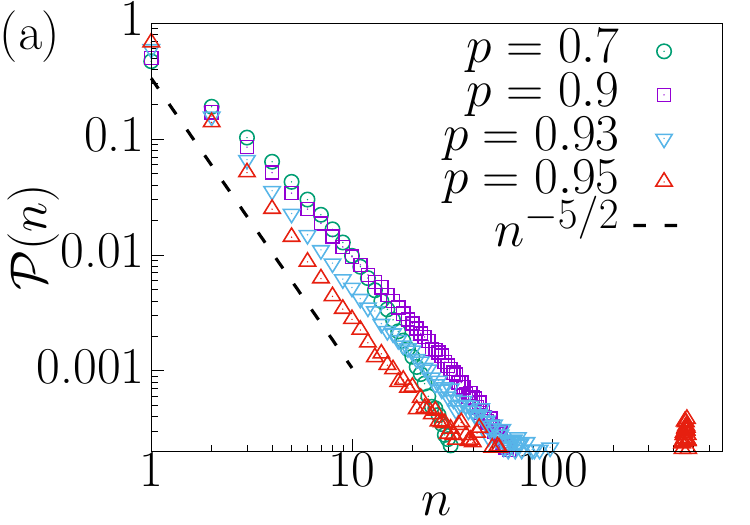}
\includegraphics[width=4.27cm,angle=0]{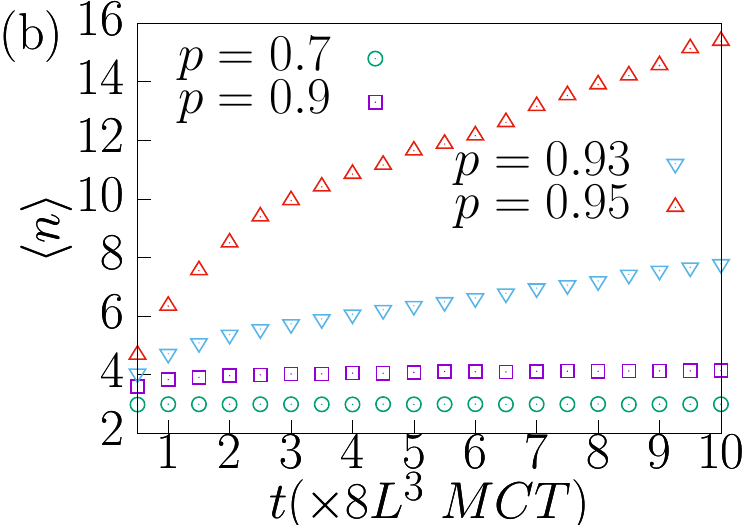}
\caption{Characterization of clustering transition. Panel (a): Plot of cluster size distribution ${\cal P}(n)$ with different cell-cell interaction rates $p$ for fixed global density $\rho=0.2$, other parameters are as in Fig.~\ref{cluster}. We observe a hump in the distribution at higher value of $p$ indicating a clustering transition while the particles moving around the cluster contribute in the power law distribution. Panel (b): Plot showing mean cluster size $\langle n \rangle$ as a function of time. The growth of $\langle n \rangle$ with time signifies the merging of small cluster into larger ones over time. Interestingly, for higher values of $p$, the time to reach the steady state is higher.}
\label{cluster-dist}
\end{center}
\end{figure}

\subsection{Clustering transition}

As we already know that clustering is a natural outcome in the system of pili-driven bacterial colony \cite{Tainer2004, Mattick2002, Maier_PRL2010, SheetzMP_Nature2000, Berg_PNAS2001, Maier2019}, we would like to verify it in the context of the proposed model. Additionally, we are interested in the influence of the key parameters responsible for the underlying dynamics on the emergence of clusters in our lattice model.
Cluster formation within the model arises due to the pili-mediated dipole attraction force which is proportional to the number of bound pili pairs between neighboring particles. Interestingly, the number of bound pairs varies with the gap in between two adjacent particles.  The initial vacancy imbalance in the two opposite directions (say, for example, left and right) of a particle created by a sudden system fluctuation results in an imbalance of attractive force in the same two opposite directions. This force imbalance magnifies further the distance difference and thus finally leads to a cluster formation. 
In the high density limit, however, this mechanism breaks down. In that limit, the mean gap between the particles becomes shorter compared to the interaction length. As a result, on average the number of bound pili-pairs for any particle is similar on both (left/right or up/down) sides. So we do not expect a clustering transition in the high density regime.

Therefore, we consider a fixed low density $\rho=N/L^2=0.2$ and visualise the cluster formation with time for a parameter regime where the cell-cell interaction rate $p=0.95$ is high. The other parameters we set at $N_p=20$, $l_0=2\ell$ and $T_0=10$ MCT. Here, $\ell$ is the lattice spacing and the time unit for the simulation of this model is defined as Monte Carlo Time (MCT), during which each particle, on average, has one opportunity to jump. In Fig.~\ref{cluster}(a), we see a gradual formation of microcolonies over time, which eventually merge to form a single large cluster surrounded by a less dense gas-like phase of particles. We also notice that it takes a significantly long time ($\sim L^4$ MCT) to reach this state of a single cluster. In panel (b) of Fig.~\ref{cluster}, we show the dependence of clustering on the parameter $p$ at time $t=L^4$ MCT while we keep the other parameters fixed as before. We observe, as expected, the clustering to increases with increasing values of $p$.

To quantify the influence of the cell-cell interaction parameter $p$ on clustering, we compute the cluster size distribution. Here, a cluster is defined as a contiguous group of linked particles connected via nearest neighbors (particles sharing an edge) and next-nearest neighbors (particles sharing a corner). In panel (a) of Fig.~\ref{cluster-dist}, we observe that the cluster size attains  a power law distribution for the surrounding gas-like phase with a condensation-like hump at long tail signifying the presence of a large cluster. Similar hump is observed also in experiments of clustering \textit{N. gonorrhoeae} bacteria \cite{Zaburdaev2015}. The reason behind the exponent of the power law $\sim -5/2$ could be an excellent open question, however, appearance of this exponent is not uncommon in the context of condensation transition \cite{Barma-model, Chakraborti_PRE2021}. In panel (b), we show the change in mean cluster size over time for different $p$. Notably, the mean cluster size is always higher for higher $p$ and it takes longer time to reach the steady state. Additionally, the growth of mean cluster size with time also indicates the coalescence of smaller clusters into larger ones as was also observed in experiments \cite{Zaburdaev2015}.

In the following section, we derive a hydrodynamic description of the system and discuss the role of the hydrodynamic transport coefficients in this clustering transition.

\section{Derivation of hydrodynamics}
\label{sec:hydro}

Our two-dimensional model with hard core occupancy can be described with occupation index of $\{i,j \}$-th site $\eta_{i,j}=1$ if the site is occupied or $\eta_{i,j}=0$ otherwise. The equilibrium limit of this model is when cell-cell interaction rate $p=0$ \cite{Chakraborti-PRR2024}. In this limit, each particle jumps to any of the four direction with equal rate $1/4$. The system in this limit is known as the simple symmetric exclusion process (SSEP) \cite{Derrida_PRL2004} which is essentially a `gradient type' process \cite{Krapivsky_PRE2014, supplement}. In appendix \cite{supplement}, we show that the $x$ and $y$ components of current vector in SSEP are independent of each other. Isotropic movements of particles and absence of cross linking between pili in orthogonal directions in our original model also allow for independence of $x$ and $y$ components of current vector when $p \neq 0$. Thus, we can derive hydrodynamics for any row or column of the 2D system which effectively will represent the hydrodynamics of the entire system. A comparison between 1D theoretical and 2D simulation results can establish the validity of this argument which we show below. It is worth noting that our previous study \cite{Chakraborti-PRR2024} examined a one-dimensional version of this system with semi-analytical results. In contrast, the present work derives fully analytical expressions for transport coefficients by obtaining analytical forms of microscopic jump weights and steady-state probability distributions. Additionally, the analysis extends to two dimensions, demonstrating the generalizability and robustness of our theoretical framework.

However, even in one-dimension, the analytical treatment of exclusion models with non-trivial interactions is challenging, particularly for calculations of correlations. This can be overcome by an exact mapping of the exclusion model (EM) to an unbounded mass transfer model (UMTM) which we discuss next \cite{Chakraborti-PRR2024}.

\subsection{Mapping to unbounded model}

In one-dimension, the exclusion model (EM) can be exactly one-to-one mapped \cite{Evans_JPA2005} to an unbounded mass transfer model (UMTM) which simplifies analytical calculation of correlators which are essential for formulating fluctuating hydrodynamics. 
We illustrate this mapping for the last row of EM in Fig.~\ref{Schem1}(b) with $N_1$ number of green particles to a corresponding UMTM presented in Fig.~\ref{Schem1}(d) (see also \cite{Chakraborti-PRR2024}). Later, during derivation of hydrodynamics we will take average over all rows and over many realisations. Note that, on average, the particle density $N_1/L$ in any row is equivalent to the particle density $\rho=N/L^2$ of the whole 2D system. In the last row of EM in Fig.~\ref{Schem1}(b), for every $i$-th (green) particle we construct a (green) lattice site labelled by $i$ in the UMTM in Fig.~\ref{Schem1}(d) resulting in a ring of $\tilde{L} = N_1$ sites. For $m_i$ vacant lattice sites in between $i$-th and $(i+1)$-th particles in the EM we put (red) mass $m_i$ at $i$-th (green) site of the UMTM, so that the total mass in the UMTM is $\tilde{N} = \sum_i m_i=L-N_1$. Thus, the mass density $\tilde{\rho}$ in the UMTM is linked to EM density $\rho$ via:
\be 
\tilde{\rho} = \tilde{N} / \tilde{L} = 1/\rho -1. 
\label{mapping}
\ee
Note that the movement of mass in UMTM is in the opposite direction of the particle movement in the EM. 
We next derive hydrodynamics and calculate transport coefficients in this 1D ring of UMTP and then use a reverse mapping to recover results in the original EM setting which is still valid for the 2D system.

\subsection{Determination of microscopic jump weights}

The first step for the derivation of hydrodynamics is the calculation of probability weight of microscopic jumps performed by the particles according to cell-substrate ($q$) and cell-cell ($p=1-q$) interactions. For example, in EM, the condition of hopping of $i$-th particle to left due to cell-substrate interaction is very simple -- at least one vacancy at the left side of $i$-th particle. This leads to the condition for the single mass transfer from site $(i-1)$ to the right direction in UMTM becoming the $(i-1)$-th site should be occupied by at least one unit of mass. We denote this condition as well as the probability weight for this mass transfer with the occupation indicator $a_{i-1}=1-\delta_{m_{i-1},0}$. The hopping of the same particle due to the cell-cell interaction depends, in addition to the occupation indicator, on the number of bound pili pairs in between $(i-1)$-th and $i$-th particles. The number of bound-pairs is proportional to the probability measure that the total length of two pili from two interacting neighboring particles is greater than or equal to the distance (equivalently mass in UMTM) $m_{i-1}$ between the particles. Below we compute this probability $v(m)$ for any distance (mass) $m$. 

\begin{figure}
\begin{center}
\includegraphics[width=4.27cm,angle=0]{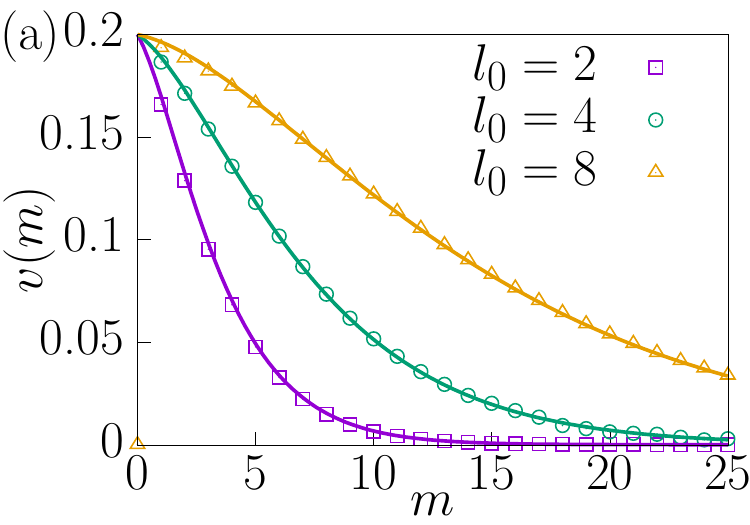}
\includegraphics[width=4.27cm,angle=0]{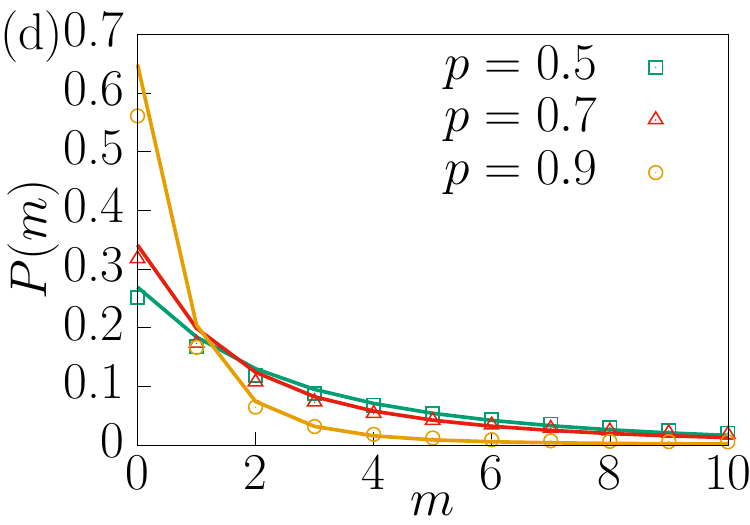}
\includegraphics[width=4.27cm,angle=0]{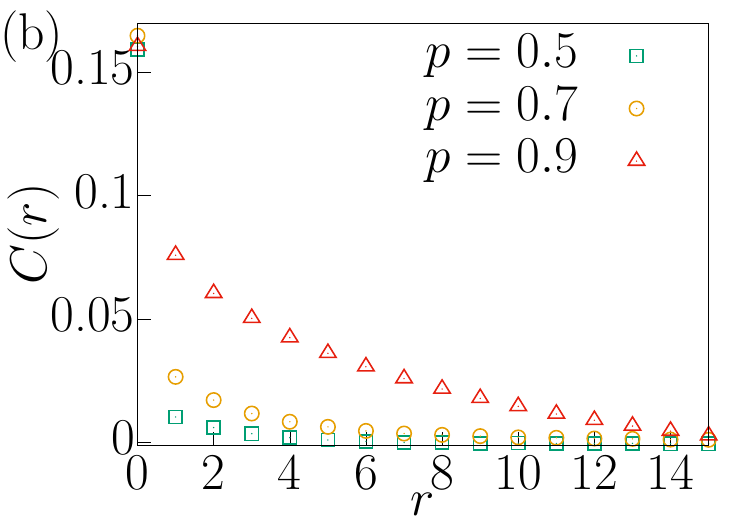}
\includegraphics[width=4.27cm,angle=0]{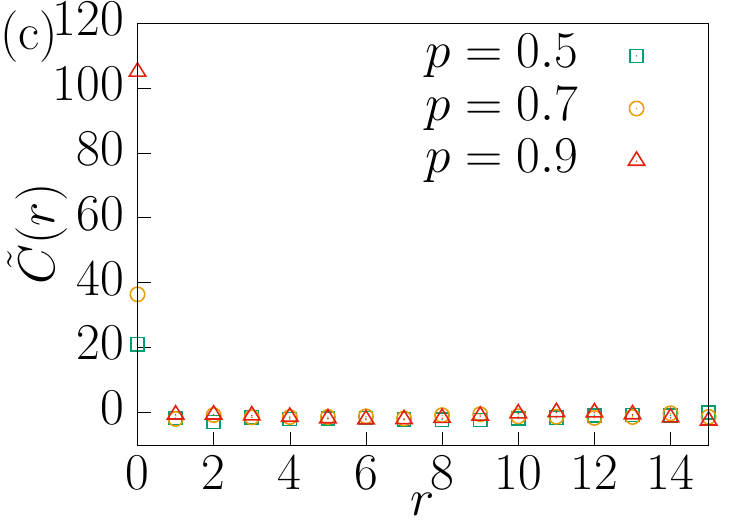}
\caption{Mass-dependent jump weights and distributions in UMTM, aiding transport coefficient analysis. Panel (a): Plot of stochastic weight $v(m)$ for hopping due to cell-cell interaction with mean gap-size $m$ within a row or column in EM for different mean pili lengths $l_0$. These results do not depend on $\rho$ and $p$, while other parameters are as in Fig.~\ref{cluster}. Symbols are the simulation results and the lines are given by formula Eq.~\eqref{eq-v_m}. Panel (b): Plot of density correlation $C(r)=\left\langle\eta_0 \eta_{r}\right\rangle - \rho^2$ in EM with distance $r$ for global density $\rho=0.2$. We can notice a long-range correlation of densities at higher cell-cell interaction weight $p$. Panel (c):  Density correlation $\tilde{C}(r)=\langle m_0m_r \rangle - \tilde{\rho}^2$ in UMTM with distance $r$ for global density $\tilde{\rho}=4$. Interestingly, in this mapping the correlation vanishes beyond distance $r=0$ for any $p$. Panel (d): Represented by symbols, the single site mass distribution $P(m)$ for different $p$ from the simulation of UMTM at density $\tilde{\rho}=4$. Lines are the corresponding analytical results given by Eq.~\eqref{P_m}.}
\label{correlation}
\end{center}
\end{figure}

In our model, the pili length $l$ is distributed exponentially with mean length $l_0$ as given by
\be
\phi(l)=\frac{1}{l_0} \exp{\left(-\frac{l}{l_0} \right)},
\label{l-dist}
\ee
and it remains unchanged until its disappearance. Consider a condition when the total length $l_1+l_2$ of two pili of random lengths $l_1$ and $l_2$ is equal to $r$. The probability $\tilde{\phi}(r)$ of this condition can be obtained using a convolution function $\phi(l_1)$  as given by:
\be
\tilde{\phi}(r)= \int_0^r \phi(l_1) \phi(r-l_1) dl_1.
\label{convol}
\ee
After inserting Eq.~\eqref{l-dist} in Eq.~\eqref{convol} we obtain:
\begin{eqnarray}
\tilde{\phi}(r) = \frac{r}{l_0^2} \exp{\left(-\frac{r}{l_0}\right)}.
\end{eqnarray}
When the sum of the length of two pili of two adjacent particles is greater than or equal to the distance $m$ between the particles then those pili will form a bound pair. In the simplest setting, the probability of forming a pili-pair
\begin{eqnarray}
\tilde{v}(r \ge m) = \int_m^\infty \tilde{\phi}(r) ~dr = \left( 1+\frac{m}{l_0} \right) \exp \left( -\frac{m}{l_0} \right). 
\end{eqnarray}
This probability provides the basic shape of the hopping weight due to cell-cell interaction. However, in addition to this probability, the number of bound-pairs depends on other factors such as mean pili lifetime $T_0$, pili number per particle $N_p$ and stochastic pili binding rate. We empirically incorporate the role of these factors through two fitting parameters $s_1$ and $s_2$ as:
\be
v(m) = \frac{s_1}{2}\left( 1+s_2\frac{m}{l_0} \right) \exp \left( -\frac{m}{l_0} \right).
\label{eq-v_m} 
\ee
From simulation of 2D EM, we compute the scaled mean number of bound-pairs $v(m)=\langle n \rangle/(2N_p)$ for distance (in both $x$ and $y$ direction) between two particles $m$ and plot them in Fig.~\ref{correlation}(a) with symbols for different mean pili lengths $l_0=2$, 4 and 8 while we keep $T_0=10$~MCT, $N_p=20$, $\rho=0.5$, and $p=0.7$. The lines are given by Eq.~\eqref{eq-v_m} with $s_1=0.4$ and $s_2(l_0=2)=0.8$, $s_2(l_0=4)=0.85$ and $s_2(l_0=8)=0.9$. We further check (not shown) that $v(m)$ does not depend on density $\rho$ and cell-cell interaction rate $p$. Throughout this work we consider $T_0=10$~MCT, $N_p=20$ and $l_0=2$, therefore, we set $s_1=0.4$ and $s_2=0.8$ for the rest of this paper.

The expression for jump weight $v(m)$ given by Eq.~\eqref{eq-v_m} is the key starting point allowing to derive an exact analytical expression for the transport coefficients, a feature absent in our earlier work \cite{Chakraborti-PRR2024}.

\subsection{Derivation of bulk-diffusivity}
\label{Sec:diffusivity}

Since we already know the microscopic mass transfer rates related to cell-substrate and cell-cell interaction, we can now put forward the master equation for mass $m_i$ by considering all possible ways of transferring a mass between the site $i$ and its neighboring sites $j=\{i-1,i+1\}$ and the associated jump weights $c_{i j}$. Thus, for UMTM, the continuous time evolution of mass $m_i(t)$ at site $i$ and at time $t$ in an infinitesimal time interval $dt$ is given by 
\begin{eqnarray}
m_i(t+dt) =
\left\{
\begin{array}{ll}
m_i(t) - 1            & {\rm prob.}~ c_{i,i-1} dt/4, \\
m_i(t) - 1            & {\rm prob.}~ c_{i,i+1} dt/4, \\
m_i(t) + 1            & {\rm prob.}~ c_{i-1,i} dt/4, \\
m_i(t) + 1            & {\rm prob.}~ c_{i+1,i} dt/4, \\
m_i(t)                & {\rm prob.}~ 1-\Sigma dt,
\end{array}
\right.
\label{unbounded-unbiased}
\end{eqnarray}
with $\Sigma=\left(c_{i,i-1}+c_{i,i+1}+c_{i-1,i}+c_{i+1,i}\right)/4$. The explicit expressions of the jump weights are: $c_{i,i-1}=qa_i+pa_iv(m^R_i)$, $c_{i,i+1}=qa_i+pa_iv(m^L_{i+1})$, $c_{i-1,i}=qa_{i-1}+pa_{i-1}v(m^L_i)$ and $c_{i+1,i}=qa_{i+1}+pa_{i+1}v(m^R_{i+1})$ with $a_i=1-\delta_{m_i,0}$ is the probability for a site $i$ to be occupied. 
Note that in EM, two neighboring particles share the same number of vacancies between them implying $m^R_i=m^L_{i+1}$ which for the UMTM system results in jump weights depending only on the initial site $i$ thus rendering it a zero-range-process \cite{Evans_JPA2005}. Here, all lattice indices are associated with a length-scale which is the lattice constant $\ell$ which we set to unity for simplicity and the time $t$ is measured in MCT. We next derive the discrete dynamical equation for the local mass density $\tilde{\rho}_i(t)=\langle m_i(t) \rangle$  and arrive at \cite{supplement}:
\begin{eqnarray}
\label{gradient1}
\partial_t \tilde{\rho}_i(t) = \frac{1}{4}\left( G_{i-1}+G_{i+1}-2G_i \right), \\
\label{g_formula}
\mbox{with~~} G_i = \langle g_i \rangle = \langle q a_i + pa_iv(m_i)\rangle.
\end{eqnarray}
The local current in Eq.~\eqref{gradient1} having a discrete gradient structure of the local observable $G_i$ renders our system of a {\it gradient type} \cite{Krapivsky_PRE2014}. This allows us to write down the time-evolution equation for the coarse-grained density field $\tilde{\rho}(x,\tau)$ in the diffusive scaling limit $i \rightarrow x = i/\tilde{L}$ and $t \rightarrow \tau = t/\tilde{L}^2$ after neglecting higher order terms ${\cal O}(1/\tilde{L}^3)$ for large system size $\tilde{L}$ as \cite{supplement}:
\begin{equation}
\frac{\partial \tilde{\rho}(x,\tau)}{\partial \tau} =  - \frac{\partial}{\partial x} \left( -D(\tilde{\rho}) \frac{\partial \tilde{\rho}}{\partial x} \right).
\label{diffusion}
\end{equation}
Here the hydrodynamic diffusive current $J_D(\tilde{\rho})=-D(\tilde{\rho}) \partial_x \tilde{\rho}$ is the mean of the current through a coarse-grained volume in the presence of a density gradient $\partial_x \tilde{\rho}$. The proportionality coefficient of this mean hydrodynamic current is given by the transport coefficient of bulk-diffusivity $D(\tilde{\rho})=[\partial G(\tilde{\rho})/\partial \tilde{\rho}]/4$. It is important to note that the bulk-diffusivity $D(\rho)$ differs from the self-diffusivity of a tagged particle \cite{Alexander_PRB1978, Krapivsky_JSP2015, Krapivsky_PRL2014} which is calculated from the mean-squared displacement (MSD) of a tagged particle as a function of time.  In a many-body system of interacting particles these diffusivities do not necessarily coincide. 
So far unknown $G(\tilde{\rho})$ is the local coarse-grained observable which has all information about microscopic interactions and thus in general also depends on other parameters $p$, $T_0$, $N_p$ and $l_0$. We will determine $G(\tilde{\rho})$ later and will use it to obtain an analytical expression for bulk-diffusivity.

The hydrodynamic current, computed within a coarse-grained volume and over a long period of time, is a stochastic quantity. Therefore, the mean hydrodynamic current, as presented in deterministic hydrodynamic equation \eqref{diffusion}, describes the `typical' coarse-grained trajectories of the system. The behavior of the `atypical' trajectories is reflected in the fluctuation of the hydrodynamic current which can bring forward new transport coefficients such as conductivity. This essentially requires the derivation of fluctuating hydrodynamics which we do next.

\subsection{Derivation of conductivity}

One useful tool to derive fluctuating hydrodynamics is the recently developed macroscopic fluctuation theory \cite{Bertini_PRL2001, Bertini_RMP2015, Derrida_JSM2007} applicable to stochastic diffusive systems having Markov property where the total number of particles is conserved. In last few decades, MFT has served to develop fluctuating hydrodynamics in several boundary driven and other lattice gas systems \cite{Bertini_RMP2015, Derrida_JSM2007, Derrida_PRL2004, Derrida_JSP2009, Derrida_JSM2009, Krapivsky_PRE2012, Krapivsky_PRL2014}, non-equilibrium mass transport models \cite{Das_PRE2017, Chakraborti_PRE2021} as well as in some inherently out-of-equilibrium models of active matter \cite{Chakraborty_PRE2020, Rahul_PRE2020, Agranov_SPP2023, Dandekar_JSM2023, Tanmoy2023}. 
MFT suggests that for a process of gradient type, a fluctuating hydrodynamics equation can be written in the form of the continuity equation \cite{Bertini_PRL2001, Bertini_RMP2015, Derrida_JSM2007, Chakraborti-PRR2024}:
\begin{equation}
\frac{\partial \tilde{\rho}(x,\tau)}{\partial \tau} =  - \frac{\partial}{\partial x} \left( -D(\tilde{\rho}) \frac{\partial \tilde{\rho}}{\partial x} +\sqrt{\chi(\tilde{\rho})} \eta(x,\tau) \right),
\label{fluc-hydro}
\end{equation}
where conductivity $\chi(\tilde{\rho})$ is another density dependent transport coefficient of interest for us and $\eta(x,\tau)$ is a coarse-grained noise term -- white but not necessarily Gaussian in general.
This conductivity contributes in the hydrodynamic drift current as a response to a random force originated due to microscopic active interactions. This random current, however, vanishes when averaged over time or realizations leaving only the typical current $J_D(\tilde{\rho})$. Therefore, as seen from Eq.~\eqref{fluc-hydro}, a straightforward way to determine $\chi(\tilde{\rho})$ is to calculate the variance of the coarse-grained current. 

In addition to that, according to the Ohm's law, the conductivity is the proportionality coefficients of the drift current in presence of an external biasing force. This external force in case of biological system could be a gravitational force when a substrate on which cells move is inclined relative to the horizontal orientation. Since this external force, unlike the internal random force, is a deterministic one, the contribution to the mean hydrodynamic current does not vanish. Using MFT, this additional drift current can be obtained analytically if the external bias is small enough such that the interaction weights due to this force obey a local detailed balance condition, does not generate an additional non-equilibrium contribution to the system. By applying a small constant biasing force field $F$, for example towards the right direction, the mass transfer weights in the biased system $c^F_{i j}$ are obtained by exponentially weighting the original rates $c_{i j}$ of Eq.~\eqref{unbounded-unbiased} with a proper normalisation as:
\begin{eqnarray}
c^F_{i j} =\frac{ 2 c_{i j} \exp\left[ \frac{\beta}{2} \hat{m}_{i j}F(j-i) \ell \right]}{\exp\left[ \frac{\beta}{2} \hat{m}_{i j}F(j-i) \ell \right] + \exp\left[ \frac{\beta}{2} \hat{m}_{j i}F(i-j)\ell \right]}.
\label{bias-rate}
\end{eqnarray}
The argument within the exponential in the RHS is an extra energy cost for transferring mass $\hat{m}_{i j}=1$ from site $i$ to $j=i\pm 1$ with $\beta=1$ and $\ell$ being the lattice spacing. Thus for small $F$, Eq.~\eqref{bias-rate} becomes $c^F_{i j} =c_{i j} \left[1+ F(j-i) \ell /2 \right]+{\cal O}(F^2)$ up to a leading order term in $F$. By incorporating the modified rates and  going to the diffusive scaling limit we obtain the hydrodynamic equation for the typical coarse-grained trajectories in the biased system as \cite{supplement}:
\begin{equation}
\frac{\partial \tilde{\rho}(x,\tau)}{\partial \tau} =  - \frac{\partial}{\partial x} \left( -D(\tilde{\rho}) \frac{\partial \tilde{\rho}}{\partial x} + \chi(\tilde{\rho})F \right),
\label{drift-diffusion}
\end{equation}
with the two transport coefficients
\begin{equation}
D(\tilde{\rho})=\frac{1}{4}\frac{\partial G(\tilde{\rho})}{\partial \tilde{\rho}} ~~\mbox{and}~~ \chi(\tilde{\rho})=\frac{1}{4}G(\tilde{\rho}).
\label{transport_unbounded}
\end{equation}

The only remaining part to achieve the transport coefficients analytically is to determine the unknown local observable $G(\tilde{\rho})$ which is an outcome of the all possible microscopic interactions. In our earlier study of 1D system \cite{Chakraborti-PRR2024}, we obtained $G(\tilde{\rho})$ by computing the mean occupation $\langle a_i \rangle$ and mean number of bound-pair of occupied cite $\langle a_i v(m_i) \rangle$ directly from microscopic simulations. In this work, however, we overcome this step by analytically calculating the steady state mass distribution which is essential for performing the averages associated with $G(\tilde{\rho})$ as in Eq.~\eqref{g_formula} -- a task we perform next.

\subsection{Analytical expression of $G(\tilde{\rho})$}
\label{Sec:analytic}

To obtain $G(\tilde{\rho})=\langle g \rangle(\tilde{\rho})$ we rely on \textit{local steady state} assumption which states that the local observable $\langle g_i \rangle \equiv G_i(\tilde{\rho}_{\mbox{loc}})$ which depends on the local density $\tilde{\rho}_{\mbox{loc}}$ can be described by the global behavior of the system at steady state with global density $\tilde{\rho}=\tilde{\rho}_{\mbox{loc}}$. In terms of probability, the local mass distribution $P_{\mbox{loc}}(m\mid \tilde{\rho}_{\mbox{loc}})$ can be replaced by the steady-state single-site mass distribution $P(m \mid \tilde{\rho})$ calculated at global density $\tilde{\rho}=\tilde{\rho}_{\mbox{loc}}$. In the rest of the paper, we use $P(m)$ instead of $P(m \mid \tilde{\rho})$ for simplicity in the notation.

One advantage of dealing with UMTM is that, in most cases, the correlations disappears beyond the nearest neighbor which is not always true for the EM version of the same system. This can be seen in our system also as presented in Fig.~\ref{correlation}. In panel (b), we see the density correlation in EM $C(r)=\langle \eta_0 \eta_r \rangle - \rho^2$ is non-zero over a significantly long range. This range increases as we increase the cell-cell interaction rate $p$. Interestingly, in panel (c), the mass-mass correlation $\tilde{C}(r)=\langle m_0 m_r \rangle - \tilde{\rho}^2$ vanishes rapidly when $r\neq0$. This observation allows us to assume that the steady state mass distribution in UMTM is a factorised measure for this zero range process \cite{Evans_JPA2005, Godreche2003}. This means that the steady state probability ${\cal P}(\{ m_i \})$ of finding the system in a configuration $\{ m_i \} = m_1, m_2, \dots , m_{\tilde{L}}$ is given by a product of factors $f(m_i)$ -- one factor for each site of the system -- i.e.
\be
{\cal P}(\{m_i \})= Z_{\tilde{N},\tilde{L}}^{-1} \prod_{i=1}^{\tilde{L}} f(m_i),
\label{factor}
\ee
with the normalisation constant
\be
Z_{\tilde{N},\tilde{L}} = \sum_{\{ m_i \}} \prod_{i=1}^{\tilde{L}} f(m_i) ~\delta \left( \sum_{i=1}^{\tilde{L}} m_i - \tilde{N} \right),
\ee
where the $\delta$-function ensures that we only include those configurations which contain $\tilde{N}$ particles in the sum.

We focus here on the probability $P(m)={\cal P}(m_1=m)$ of finding $m$ masses in the generic site $i = 1$. Rest of the $\tilde{L}-1$ sites will then act as a particle reservoir with chemical potential $\mu$. Thus the mass distribution at a single site can be written as:
\be
P(m)=\frac{f(m)z^m}{F(z)},
\label{P_m}
\ee
with $z=e^\mu$ and $F(z)=\sum_{m=0}^\infty f(m)z^m$ is the normalisation constant. As a consequence of the factorised steady state condition we can write \cite{Godreche2003}
\be
f(m)=\frac{1}{u(1)u(2)u(3)\dots u(m)}
\label{eq-fact},
\ee
with $u(m)=q+pv(m)$ and $f(0)=1$. To obtain $z(\tilde{\rho})$ we need to use the condition:
\be
\tilde{\rho}=\langle m \rangle = \frac{\sum_{m=0}^\infty mf(m)z^m}{F(z)}=z\frac{F^\prime(z)}{F(z)}.
\label{z_rho}
\ee

We use Eqs.~\eqref{eq-v_m} and \eqref{eq-fact} to compute $F(z)$ numerically within the limit of $z$ such that the series $\sum_{m=0}^\infty f(m)z^m$ converges. Then using Eq.~\eqref{z_rho} we determine $z(\tilde{\rho})$ and put that in Eq.~\eqref{P_m} to obtain $P(m)$ for different values of density $\tilde{\rho}$ and activity $p$. We test our analytical finding with the numerical simulation of UMTM at global density $\tilde{\rho}=4$. In Fig.~\ref{correlation}(d), the symbols represent the single site mass distribution $P(m)$ for different activity $p$ obtained from simulation whereas the corresponding lines are the analytical results given by Eq.~\eqref{P_m}. The good agreement between theory and simulation allows us to use this $P(m)$ to perform the average in the calculation of the essential observable $G(\tilde{\rho})=\langle g \rangle(\tilde{\rho})$.

To proceed further, we calculate the rate of change $d\langle m_i^2 \rangle /dt$ of the second moment of mass $m_i$ from Eq.~\eqref{unbounded-unbiased}:
\begin{widetext}
\begin{eqnarray}
\frac{d\langle m_i^2 \rangle}{dt} = \frac{\langle m_i^2(t+dt) \rangle - \langle m_i^2(t) \rangle}{dt} = \langle (m_i-1)^2 g_i \rangle + \frac{1}{2}\langle (m_i+1)^2 (g_{i+1}+g_{i-1}) \rangle - \langle m_i^2 [(g_{i+1}+g_{i-1})/2+g_i] \rangle.
\end{eqnarray}
\end{widetext}
The vanishing correlation in Fig.~\ref{correlation}(c) when $r>0$ makes our calculation simple. For any two observables ${\cal B}$ and ${\cal C}$, we can now write $\langle {\cal B}_i {\cal C}_j \rangle = \langle {\cal B}_i\rangle \langle {\cal C}_j \rangle$ if $i \neq j$. At steady state $d\langle m_i^2 \rangle /dt$ is zero which lead us to:
\begin{eqnarray} \nn
-2\langle m_ig_i \rangle + \langle g_i \rangle -2\langle m_ig_i \rangle +\langle g_i \rangle + 2\langle m_i \rangle \langle g_{i+1} \rangle \\
 + \langle g_{i+1} \rangle +2\langle m_i \rangle \langle g_{i-1} \rangle +\langle g_{i-1} \rangle = 0.
\label{observable}
\end{eqnarray}
We notice that the clusters at the steady state are not fixed in there position over the time. Therefore, within a large coarse-grained volume and in a span of long time the local observables effectively become homogeneous and thus we can remove the local index $i$ from everywhere and obtain
\begin{eqnarray} 
\langle g \rangle =\frac{\langle mg \rangle}{1+\tilde{\rho}}.
\label{gi}
\end{eqnarray}
Now, we use the definition of occupancy $a=1-\delta_{m,0}$ to simplify 
\begin{eqnarray} \nn
\langle mg \rangle &=& \langle m(1-\delta_{m,0}) [q+pv(m)] \rangle , \\
&=& \langle m[q+pv(m)] \rangle .
\end{eqnarray}
We use $P(m)$ to calculate the averages and use the relation $\tilde{\rho}=\langle m \rangle$ to obtain
\begin{eqnarray}
\langle mg \rangle = q\tilde{\rho} +p \sum_{m=0}^\infty mv(m)P(m).
\label{mg}
\end{eqnarray}
Combining Eqs.~\eqref{gi} and \eqref{mg} we finally obtain the observable
\begin{eqnarray}
G(\tilde{\rho}) = \frac{q\tilde{\rho} +p \sum_{m=0}^\infty mv(m)P(m)}{1+\tilde{\rho}}.
\label{g_rho}
\end{eqnarray}

\subsection{Reverse mapping and transport coefficients}

In Eqs.~\eqref{transport_unbounded}, we obtained the transport coefficients for the UMTM as a function of mass density $\tilde{\rho}$. To obtain them in the actual EM we first express the observable $G$ in terms of particle density $\rho$ using Eq.~\eqref{mapping} into Eq.~\eqref{g_rho}
\begin{eqnarray} 
G(\rho) = q(1-\rho) + p \rho \sum_{m=0}^\infty mv(m)P(m).
\label{G-analytic-exclusion}
\end{eqnarray}
Next, we use the reverse mapping of UMTM to EM representation \cite{Rizkallah_JSM2023} and recover the fluctuating hydrodynamics description for original EM with a similar structure as in Eq.~\eqref{fluc-hydro} with the two modified transport coefficients: 
\begin{eqnarray}
D({\rho})&=&-\frac{1}{4}\frac{\partial G({\rho})}{\partial {\rho}} ~~\mbox{and}~~ \chi({\rho})=\frac{1}{4}\rho G({\rho}).
\label{transport-EP}
\end{eqnarray}

As we discussed earlier, the independence in the movement of particles in $x$ and $y$ direction makes the components of current vector independent and thus both components are characterized by the same transport coefficients. Therefore, in two-dimension, the fluctuating (for unbiased system) and deterministic (for biased system) hydrodynamic equation respectively can be written as:
\bea
\label{unbiased-diffusion-2D}
\frac{\partial \rho(\textbf{r},\tau)}{\partial \tau} = -\bigtriangledown . \left[ -D(\rho) \bigtriangledown \rho + \sqrt{\chi(\rho)} \vec{\eta}(\textbf{r},\tau) \right], \\
\frac{\partial \rho(\textbf{r},\tau)}{\partial \tau} = -\bigtriangledown . \left[ -D(\rho) \bigtriangledown \rho + \chi(\rho) \textbf{F} \right],
\label{drift-diffusion-2D}
\eea
with final analytical expression of the two transport coefficients
\begin{eqnarray}
\label{d_rho}
D(\rho) = \frac{1}{4}\left[ q - p \frac{\partial}{\partial {\rho}} \left\{ \rho \sum_{m=0}^\infty mv(m)P(m\mid \rho) \right\} \right], \\
\chi(\rho) = \frac{1}{4}\left[ q \rho (1-\rho) + p \rho^2 \sum_{m=0}^\infty mv(m)P(m\mid \rho) \right].
\label{chi_rho}
\end{eqnarray}
The transport coefficients shown in Eqs.~\eqref{d_rho} and \eqref{chi_rho} are our main results. They, in general, depend also on cell-cell interaction parameter $p$, pili number per particle $N_p$, mean pili length $l_0$ and lifetime $T_0$. Below we compare our analytical results with the same observed from Monte Carlo simulations of a two-dimensional lattice system.

\section{Simulation results}
\label{sec:result}

\begin{figure}
\begin{center}
\includegraphics[width=4.27cm,angle=0]{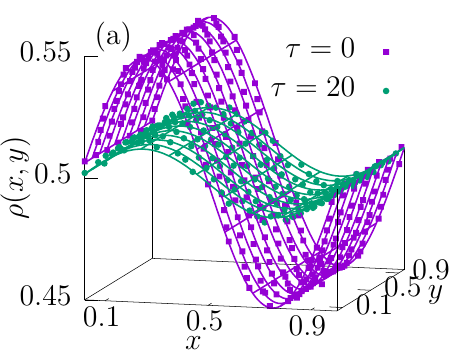}
\includegraphics[width=4.27cm,angle=0]{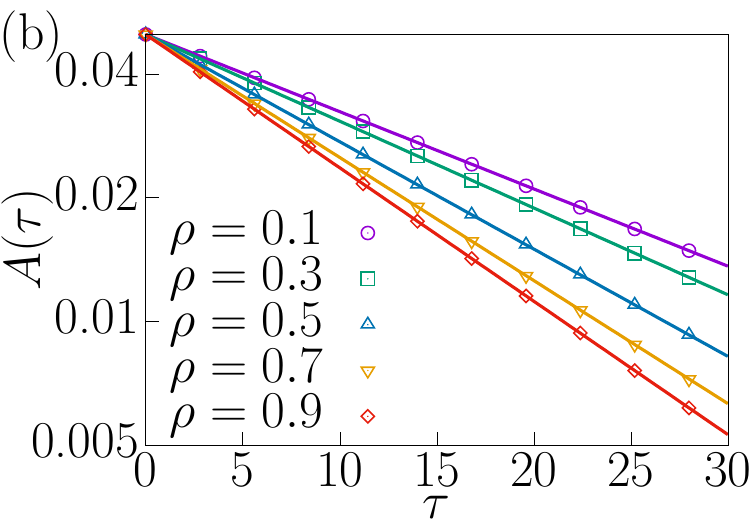}
\includegraphics[width=4.27cm,angle=0]{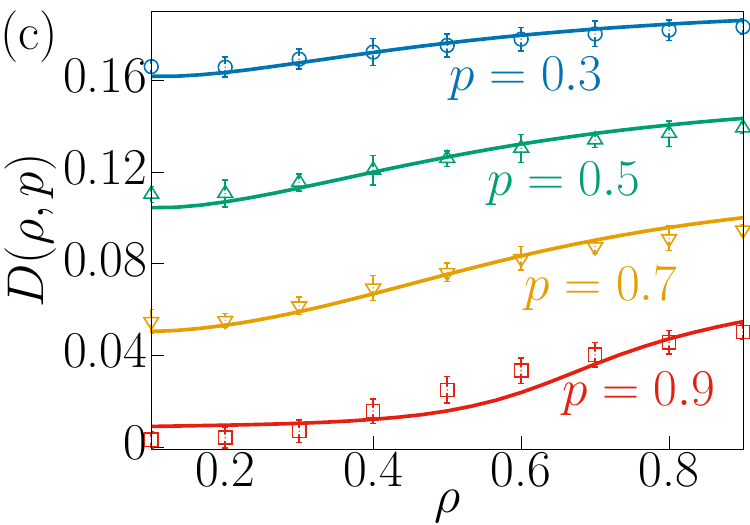}
\includegraphics[width=4.27cm,angle=0]{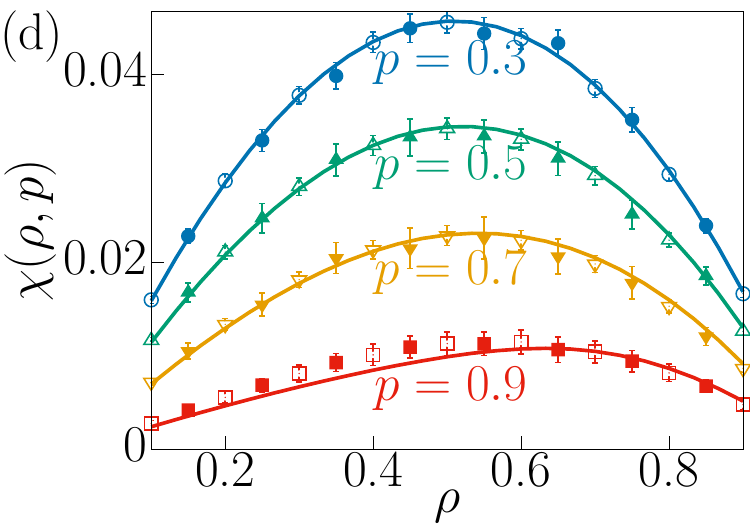}
\caption{Determination and demonstration of transport coefficients. Panel (a): Relaxation of sinusoidal perturbation with initial amplitude $A(0)=0.05$ along $x$-direction on top of a background density $\rho_0=0.5$ for $p=0.7$. The symbols are simulation data and the lines are $\rho(x,\tau)=\rho_0+A(\tau)\sin(2\pi x)$. Panel (b): Plot of $A(\tau)$ with scaled time $\tau$ in semi-log scale for $p=0.7$ for different background densities $\rho_0$ which we denote by $\rho$ due to local steady state assumption. Symbols are simulation results and lines are fitting with Eq.~\eqref{amplitude} which provides the bulk-diffusivity. Panel (c) shows (by symbols) the bulk-diffusivity $D(\rho,p)$ obtained from the exponent of Eq.~\eqref{amplitude} and compares (presented by lines) it with the corresponding analytical values calculated from Eq.~\eqref{d_rho} for different $\rho$ and $p$. Panel (d): Solid symbols represent the conductivity $\chi(\rho,p)$ numerically computed from the mean current of a biased system whereas the empty symbols are the same computed from the variance of current in an unbiased system. We compare them with analytical results in Eq.~\eqref{chi_rho} shown by lines. The error bars in panel (c) combines the propagated fitting error from panel (b) with the local fitting error specific to panel (c). In panel (d), the error bar is given by the standard error of the mean. For all other figures in the paper, the error bars are smaller than the representative symbol sizes of the data; therefore, we chose not to display error bars to avoid unnecessary visual clutter.
}
\label{fig-transport}
\end{center}
\end{figure}

In the final part of the paper, we present the results from Monte Carlo simulation of the microscopic model to compare them with the results obtained analytically. As discussed in Sec.~\ref{Sec:analytic}, local steady state is an essential criterion for our theoretical framework. A key requirement for meeting this criterion is ensuring that all observables are computed at steady state. However, as shown in Fig.~\ref{cluster-dist}, we notice that at low density limit $\rho=0.2$, and beyond $p=0.9$, where the clustering becomes more prominent, the system requires an arbitrarily long time to reach steady state. Consequently, within the limits of our computational resources, we limit our simulation studies to $p \leq 0.9$. We first focus on the verification of the bulk-diffusivity. 

To determine bulk-diffusivity from simulation, we study the relaxation of a density profile $\rho(x,0)=\rho_0+\delta\rho(x,0)$ with a sinusoidal small perturbation $\delta\rho(x,0) = A(0)\sin(2\pi x)$ on a background density $\rho_0$, $A(0) << \rho_0$. We apply the perturbation in one spatial direction ($x$), while density profile is uniform in the other direction ($y$). In accordance with the rescaling in Sec.~\ref{Sec:diffusivity} we set the lattice position $i = xL$ which gives the wave number $2\pi/L$. Similarly, the Monte Carlo time $t$ of simulation is related to the hydrodynamic rescaled time $\tau$ as $t = \tau L^2$. Since the density profile $\rho(x,\tau) = \rho_0 + A(0)\exp[-4\pi^2D(\rho_0,p)\tau]\sin(2\pi x)$ satisfies the diffusion equation Eq.~\eqref{diffusion} (by replacing $\tilde{\rho}$ by $\rho$) as well as the initial condition, we can say that it is the solution of diffusion equation at long time. Note that the solution is also a sinusoidal one with a time dependent amplitude 
\be
A(\tau)=A(0)\exp[-4\pi^2D(\rho_0,p)\tau],
\label{amplitude}
\ee
which has the information of bulk-diffusivity $D(\rho_0,p)$.

We demonstrate the sinusoidal evolution of the density profile in Fig.~\ref{fig-transport}(a) for cell-cell interaction parameter $p=0.7$ and background density $\rho_0=0.5$. We see that the profile remains sinusoidal over time with a decreasing amplitude. In Fig.~\ref{fig-transport}(b), we present the amplitude $A(\tau)$ with rescaled time $\tau$ for different background densities $\rho_0$ which shows that it decays exponentially. Now from the exponents of $A(\tau)$, as seen in Eq.~\eqref{amplitude}, we determine the bulk-diffusivity $D(\rho_0,p)$ for different $\rho_0$ and $p$. In the limit $A(0) << \rho_0$, following the local steady state condition we can replace now the global density $\rho_0$ by the local density $\rho$. In panel (c) of Fig.~\ref{fig-transport}, we plot thus calculated $D(\rho,p)$ (shown by points) and compare them with the corresponding analytical results (lines) given by Eq.~\eqref{d_rho}. We find a very good agreement between theory and simulation and observe that $D(\rho)$ decreases with increasing $p$ mirroring the clustering phenomenon. We also observe, for a fixed $p$, a lower diffusivity at low density regime where the clustering is more prominent. Note that, in the equilibrium limit $(p=0)$, the bulk-diffusivity obtained from diffusive flux is constant for all densities \cite{Derrida_PRL2004, Chakraborty_PRE2020}. However, in the same limit, the individual tagged particles move sub-diffusively with MSD $\sim t^{1/2}$. Therefore, the self-diffusivity, understood as the proportionality constant in linear dependence of the MSD as a function of time, can not be defined 
\cite{Alexander_PRB1978, Krapivsky_JSP2015, Krapivsky_PRL2014}.

Next, we calculate conductivity $\chi(\rho)$ from simulation in two ways. One way is by computing the variance of the hydrodynamic current as given by Eq.~\eqref{unbiased-diffusion-2D} originating due to the in-build random force of the system. We compute that coarse-grained current 
\be
\vec{{\cal J}}=\sum_{i=1}^L j_i^{(x)}\hat{x} +\sum_{k=1}^L j_k^{(y)}\hat{y},
\label{current2}
\ee
where $j_i^{(x)}$ and $j_k^{(y)}$ are the $x$- and $y$-component of time integrated current through horizontal bond $(i,i+1)$ and vertical bond $(k,k+1)$, respectively, during a long time $t$. Clearly, at steady state, for diffusive systems without a bias $\left \langle \vec{{\cal J}} \right \rangle=0$, however, the fluctuation of this current accounting for the atypical coarse-grained trajectories is related to the conductivity through $\chi(\rho)=\left \langle \vec{{\cal J}}^2 \right \rangle/(4Lt)$ \cite{Derrida_PRL2004, Tanmoy2023}. We show this conductivity in panel (d) of Fig.~\ref{fig-transport} by empty symbols which agree remarkably with the corresponding analytical lines calculated using Eq.~\eqref{chi_rho}. In the second case, we calculate the mean current in presence of a small bias $\textbf{F}$ with magnitude $|\textbf{F}|= F=0.1$ in the system. At steady state of the biased system, we compute mean coarse-grained current $\left \langle \vec{{\cal J}} \right \rangle /t$ over a long time $t$ and by averaging over many realisations. According to Eq.~\eqref{drift-diffusion-2D}, this current should be equal to $\chi(\rho) \textbf{F}$. The $\chi(\rho)$ calculated in this way is presented by solid symbols in panel (d) of Fig.~\ref{fig-transport} where the agreement with the lines obtained analytically is again excellent.
Similar to bulk-diffusivity, the conductivity also decreases with increasing cell-cell interaction parameter $p$ as an indication of a clustering transition. The conductivity at equilibrium limit $p=0$ is $\chi(\rho)=\rho (1-\rho)/4$ \cite{Derrida_PRL2004, Chakraborty_PRE2020}, after which as we increase $p$, an additional asymmetry appears across $\rho=0.5$.

Note that there is a mismatch between simulation and analytical results in Figs.~\ref{fig-transport}(c) and \ref{fig-transport}(d) in the high cell-cell interaction ($p = 0.9$) limit which is not due to statistical uncertainty. Rather, it arises from approximations inherent in both approaches. On the analytical side, two key assumptions underlie the results: (i) the dynamics of the 2D system are approximated by an effective 1D model, and (ii) the steady-state mass distribution is assumed to factorize [as in Eq.~\eqref{factor}], supported by the negligible density correlations seen in Fig.~\ref{correlation}(c). Similar assumptions are used in deriving Eqs.~\eqref{eq-fact} and \eqref{observable}, but they are likely to break down at high activity levels $p$, especially in 2D.
On the simulation side, finite-size effects—especially in the calculation of diffusivity—become significant. The accuracy of the sinusoidal perturbation method improves with system size and smaller perturbation amplitude. However, since both transport coefficients decrease with increasing 
$p$, the system takes longer to reach steady state, enhancing finite-size and finite-time limitations.

\section{Conclusions}
\label{conclusion}

In this study, we explore the large-scale transports in a system of cellular aggregates driven by attractive dipole forces. We consider a simple model of \textit{N. gonorrhoeae} bacterial colony interacting via pili-mediated contractile forces. 
Our model can capture the essential features of clustering transition through coalescence of smaller micro-colonies. We then explore this behavior in the context of macroscopic currents by deriving a fluctuating hydrodynamics. The first step towards it is to analytically express the stochastic jump weights -- which we use in a recently developed Macroscopic Fluctuation Theory. This theory provides us with two density-dependent hydrodynamic transport coefficients -- bulk-diffusivity and conductivity -- the proportional coefficients of the hydrodynamics currents in presence of a density gradient and an external bias, respectively. The final challenge for arriving to a complete analytical formula is calculation of observable correlators directly related to the transport coefficients. We overcome this challenge by using a mapping from exclusion model to unbounded mass transport model and then obtaining the steady state mass distribution in the latter model. 

The final expressions of transport coefficients are then tested with microscopic simulations. We study the relaxation of a small sinusoidal perturbation on top of a background density to obtain the bulk-diffusivity as a function of density. The conductivity is verified in two ways. On one hand, we measure the mean hydrodynamic current in presence of an external biasing force. On the other hand, we compute the variance of the fluctuating current in absence of any bias. Both of the measurements provide the same conductivity. Both the transport coefficients -- bulk-diffusivity and conductivity agree well with the analytical expressions and show a decrease with increasing cell-cell interactions -- as a consequence of cluster formation. 
We emphasize that, despite the inherently out-of-equilibrium nature of our system \cite{Chakraborti-PRR2024}, we are able to derive explicit analytical expressions for the hydrodynamic transport coefficients in our bacterial model, which exhibits rich clustering behavior. Such results are still rare in the studies of interacting active matter, where the complexity and non-equilibrium dynamics often lead to reliance on phenomenological approaches \cite{Cates_review}.

In this study, we focus only on the variance of the fluctuating current at steady state. Detailed steady state distribution along with dynamical behavior of the current could be interesting topics to address \cite{Tanmoy2023, Dandekar_JSM2023}.
Another immediate outlook is to study the same model with allowed cross-linking among the pili growing in horizontal and vertical directions. Clearly, the breakdown of the independence of currents in $x$- and $y$-direction could give rise to exciting characteristics, however, analytical treatment would be challenging. Moreover, we know, for equilibrium SSEP, that the self-diffusion of tagged particle is completely different from the hydrodynamic bulk-diffusivity. It will be interesting to check the self-diffusivity and tagged particle current for our non-equilibrium model.

Another important direction for future work is to perform a linear instability analysis on the analytical form Eq.~\eqref{unbiased-diffusion-2D} of $D(\rho)$ to assess whether the homogeneous phase is truly unstable. However, since $D(\rho)$ depends on the mass distribution  $P(m|\rho)$, which lacks a closed-form and must be evaluated numerically through the fugacity, an analytical treatment is intractable. A numerical approach may be feasible and offers a promising way for future investigation.

In this work, we provide a framework to derive fluctuating hydrodynamics which also should work for other systems interacting via contractile forces such as tumor spheroids \cite{Wolfgang_Review1987, Friedl_NatureReview2003, Fabry_PLoS2012, Fabry_eLife2020} or neuronal organoids \cite{Karow_Nature2018, Karow_2021, Lancaster_Nature2013, Pasca_Nature2018}. However, this framework can be extended for a general setting where the detail reason for the microscopic interaction is not available. In most experiments, the microscopic description available is the trajectories. Therefore, formulation of a generalised framework of deriving fluctuating hydrodynamics irrespective of microscopic differences could be an outstanding future plan.

\section*{ACKNOWLEDGMENTS}

We thank Arghya Das and Tanmoy Chakraborty for helpful discussions. We also thank N. Biais for providing image for Fig. \ref{Schem1}(a). V.Z. acknowledges financial support by the Deutsche Forschungsgemeinschaft (DFG, German Research Foundation) Project No. 460333672--CRC 1540 Exploring Brain Mechanics. Numerical simulations were performed on HPC cluster of NHR@FAU.

\bibliography{proposal}

\begin{thebibliography}{67}%
\makeatletter
\providecommand \@ifxundefined [1]{%
 \@ifx{#1\undefined}
}%
\providecommand \@ifnum [1]{%
 \ifnum #1\expandafter \@firstoftwo
 \else \expandafter \@secondoftwo
 \fi
}%
\providecommand \@ifx [1]{%
 \ifx #1\expandafter \@firstoftwo
 \else \expandafter \@secondoftwo
 \fi
}%
\providecommand \natexlab [1]{#1}%
\providecommand \enquote  [1]{``#1''}%
\providecommand \bibnamefont  [1]{#1}%
\providecommand \bibfnamefont [1]{#1}%
\providecommand \citenamefont [1]{#1}%
\providecommand \href@noop [0]{\@secondoftwo}%
\providecommand \href [0]{\begingroup \@sanitize@url \@href}%
\providecommand \@href[1]{\@@startlink{#1}\@@href}%
\providecommand \@@href[1]{\endgroup#1\@@endlink}%
\providecommand \@sanitize@url [0]{\catcode `\\12\catcode `\$12\catcode
  `\&12\catcode `\#12\catcode `\^12\catcode `\_12\catcode `\%12\relax}%
\providecommand \@@startlink[1]{}%
\providecommand \@@endlink[0]{}%
\providecommand \url  [0]{\begingroup\@sanitize@url \@url }%
\providecommand \@url [1]{\endgroup\@href {#1}{\urlprefix }}%
\providecommand \urlprefix  [0]{URL }%
\providecommand \Eprint [0]{\href }%
\providecommand \doibase [0]{https://doi.org/}%
\providecommand \selectlanguage [0]{\@gobble}%
\providecommand \bibinfo  [0]{\@secondoftwo}%
\providecommand \bibfield  [0]{\@secondoftwo}%
\providecommand \translation [1]{[#1]}%
\providecommand \BibitemOpen [0]{}%
\providecommand \bibitemStop [0]{}%
\providecommand \bibitemNoStop [0]{.\EOS\space}%
\providecommand \EOS [0]{\spacefactor3000\relax}%
\providecommand \BibitemShut  [1]{\csname bibitem#1\endcsname}%
\let\auto@bib@innerbib\@empty
\bibitem [{\citenamefont {Ben-Jacob}\ \emph {et~al.}(1998)\citenamefont
  {Ben-Jacob}, \citenamefont {Cohen},\ and\ \citenamefont
  {Gutnick}}]{Gutnick1998}%
  \BibitemOpen
  \bibfield  {author} {\bibinfo {author} {\bibfnamefont {E.}~\bibnamefont
  {Ben-Jacob}}, \bibinfo {author} {\bibfnamefont {I.}~\bibnamefont {Cohen}},\
  and\ \bibinfo {author} {\bibfnamefont {D.~L.}\ \bibnamefont {Gutnick}},\
  }\bibfield  {title} {\bibinfo {title} {Cooperative organization of bacterial
  colonies: From genotype to morphotype},\ }\href
  {https://doi.org/10.1146/annurev.micro.52.1.779} {\bibfield  {journal}
  {\bibinfo  {journal} {Annu. Rev. Microbiol.}\ }\textbf {\bibinfo {volume}
  {52}},\ \bibinfo {pages} {779} (\bibinfo {year} {1998})}\BibitemShut
  {NoStop}%
\bibitem [{\citenamefont {Taktikos}\ \emph {et~al.}(2015)\citenamefont
  {Taktikos}, \citenamefont {Lin}, \citenamefont {Stark}, \citenamefont
  {Biais},\ and\ \citenamefont {Zaburdaev}}]{Zaburdaev2015}%
  \BibitemOpen
  \bibfield  {author} {\bibinfo {author} {\bibfnamefont {J.}~\bibnamefont
  {Taktikos}}, \bibinfo {author} {\bibfnamefont {Y.~T.}\ \bibnamefont {Lin}},
  \bibinfo {author} {\bibfnamefont {H.}~\bibnamefont {Stark}}, \bibinfo
  {author} {\bibfnamefont {N.}~\bibnamefont {Biais}},\ and\ \bibinfo {author}
  {\bibfnamefont {V.}~\bibnamefont {Zaburdaev}},\ }\bibfield  {title} {\bibinfo
  {title} {Pili-induced clustering of n. gonorrhoeae bacteria},\ }\href
  {https://doi.org/10.1371/journal.pone.0137661} {\bibfield  {journal}
  {\bibinfo  {journal} {PLoS ONE}\ }\textbf {\bibinfo {volume} {10}},\ \bibinfo
  {pages} {e0137661} (\bibinfo {year} {2015})}\BibitemShut {NoStop}%
\bibitem [{\citenamefont {Mueller-Klieser}(1987)}]{Wolfgang_Review1987}%
  \BibitemOpen
  \bibfield  {author} {\bibinfo {author} {\bibfnamefont {W.}~\bibnamefont
  {Mueller-Klieser}},\ }\bibfield  {title} {\bibinfo {title} {Multicellular
  spheroids: A review on cellular aggregates in cancer research},\ }\href
  {https://doi.org/10.1007/BF00391431} {\bibfield  {journal} {\bibinfo
  {journal} {J. Cancer. Res. Clin. Oncol.}\ }\textbf {\bibinfo {volume}
  {113}},\ \bibinfo {pages} {101} (\bibinfo {year} {1987})}\BibitemShut
  {NoStop}%
\bibitem [{\citenamefont {Friedl}\ and\ \citenamefont
  {Wolf}(2003)}]{Friedl_NatureReview2003}%
  \BibitemOpen
  \bibfield  {author} {\bibinfo {author} {\bibfnamefont {P.}~\bibnamefont
  {Friedl}}\ and\ \bibinfo {author} {\bibfnamefont {K.}~\bibnamefont {Wolf}},\
  }\bibfield  {title} {\bibinfo {title} {Tumour-cell invasion and migration:
  diversity and escape mechanisms},\ }\href {https://doi.org/10.1038/nrc1075}
  {\bibfield  {journal} {\bibinfo  {journal} {Nat. Rev. Cancer}\ }\textbf
  {\bibinfo {volume} {3}},\ \bibinfo {pages} {362} (\bibinfo {year}
  {2003})}\BibitemShut {NoStop}%
\bibitem [{\citenamefont {Koch}\ \emph {et~al.}(2012)\citenamefont {Koch},
  \citenamefont {Munster}, \citenamefont {Bonakdar}, \citenamefont {Butler},\
  and\ \citenamefont {Fabry}}]{Fabry_PLoS2012}%
  \BibitemOpen
  \bibfield  {author} {\bibinfo {author} {\bibfnamefont {T.~M.}\ \bibnamefont
  {Koch}}, \bibinfo {author} {\bibfnamefont {S.}~\bibnamefont {Munster}},
  \bibinfo {author} {\bibfnamefont {N.}~\bibnamefont {Bonakdar}}, \bibinfo
  {author} {\bibfnamefont {J.~P.}\ \bibnamefont {Butler}},\ and\ \bibinfo
  {author} {\bibfnamefont {B.}~\bibnamefont {Fabry}},\ }\bibfield  {title}
  {\bibinfo {title} {3d traction forces in cancer cell invasion},\ }\href
  {https://doi.org/10.1371/journal.pone.0033476} {\bibfield  {journal}
  {\bibinfo  {journal} {PLoS ONE}\ }\textbf {\bibinfo {volume} {7}},\ \bibinfo
  {pages} {e33476} (\bibinfo {year} {2012})}\BibitemShut {NoStop}%
\bibitem [{\citenamefont {Mark}\ \emph {et~al.}(2020)\citenamefont {Mark} \emph
  {et~al.}}]{Fabry_eLife2020}%
  \BibitemOpen
  \bibfield  {author} {\bibinfo {author} {\bibfnamefont {C.}~\bibnamefont
  {Mark}} \emph {et~al.},\ }\bibfield  {title} {\bibinfo {title} {Collective
  forces of tumor spheroids in three-dimensional biopolymer networks},\ }\href
  {https://doi.org/10.7554/eLife.51912} {\bibfield  {journal} {\bibinfo
  {journal} {eLife}\ }\textbf {\bibinfo {volume} {9}},\ \bibinfo {pages}
  {e51912} (\bibinfo {year} {2020})}\BibitemShut {NoStop}%
\bibitem [{\citenamefont {Karow}\ \emph {et~al.}(2018)\citenamefont {Karow}
  \emph {et~al.}}]{Karow_Nature2018}%
  \BibitemOpen
  \bibfield  {author} {\bibinfo {author} {\bibfnamefont {M.}~\bibnamefont
  {Karow}} \emph {et~al.},\ }\bibfield  {title} {\bibinfo {title} {Direct
  pericyte-to-neuron reprogramming via unfolding of a neural stem cell-like
  program},\ }\href {https://doi.org/10.1038/s41593-018-0168-3} {\bibfield
  {journal} {\bibinfo  {journal} {Nat. Neurosci.}\ }\textbf {\bibinfo {volume}
  {21}},\ \bibinfo {pages} {932} (\bibinfo {year} {2018})}\BibitemShut
  {NoStop}%
\bibitem [{\citenamefont {Falk}\ \emph {et~al.}(2021)\citenamefont {Falk},
  \citenamefont {Han},\ and\ \citenamefont {Karow}}]{Karow_2021}%
  \BibitemOpen
  \bibfield  {author} {\bibinfo {author} {\bibfnamefont {S.}~\bibnamefont
  {Falk}}, \bibinfo {author} {\bibfnamefont {D.}~\bibnamefont {Han}},\ and\
  \bibinfo {author} {\bibfnamefont {M.}~\bibnamefont {Karow}},\ }\bibfield
  {title} {\bibinfo {title} {Cellular identity through the lens of direct
  lineage reprogramming},\ }\href {https://doi.org/10.1016/j.gde.2021.06.015}
  {\bibfield  {journal} {\bibinfo  {journal} {Curr. Opin. Genet. Dev.}\
  }\textbf {\bibinfo {volume} {70}},\ \bibinfo {pages} {97} (\bibinfo {year}
  {2021})}\BibitemShut {NoStop}%
\bibitem [{\citenamefont {Lancaster}\ \emph {et~al.}(2013)\citenamefont
  {Lancaster} \emph {et~al.}}]{Lancaster_Nature2013}%
  \BibitemOpen
  \bibfield  {author} {\bibinfo {author} {\bibfnamefont {M.}~\bibnamefont
  {Lancaster}} \emph {et~al.},\ }\bibfield  {title} {\bibinfo {title} {Cerebral
  organoids model human brain development and microcephaly},\ }\href
  {https://doi.org/10.1038/nature12517} {\bibfield  {journal} {\bibinfo
  {journal} {Nature}\ }\textbf {\bibinfo {volume} {501}},\ \bibinfo {pages}
  {373} (\bibinfo {year} {2013})}\BibitemShut {NoStop}%
\bibitem [{\citenamefont {Pasca}(2018)}]{Pasca_Nature2018}%
  \BibitemOpen
  \bibfield  {author} {\bibinfo {author} {\bibfnamefont {S.~P.}\ \bibnamefont
  {Pasca}},\ }\bibfield  {title} {\bibinfo {title} {The rise of
  three-dimensional human brain cultures},\ }\href
  {https://doi.org/10.1038/nature25032} {\bibfield  {journal} {\bibinfo
  {journal} {Nature}\ }\textbf {\bibinfo {volume} {553}},\ \bibinfo {pages}
  {437} (\bibinfo {year} {2018})}\BibitemShut {NoStop}%
\bibitem [{\citenamefont {P\"onisch}\ \emph {et~al.}(2017)\citenamefont
  {P\"onisch}, \citenamefont {Weber}, \citenamefont {Juckeland}, \citenamefont
  {Biais},\ and\ \citenamefont {Zaburdaev}}]{Zaburdaev2017}%
  \BibitemOpen
  \bibfield  {author} {\bibinfo {author} {\bibfnamefont {W.}~\bibnamefont
  {P\"onisch}}, \bibinfo {author} {\bibfnamefont {C.~A.}\ \bibnamefont
  {Weber}}, \bibinfo {author} {\bibfnamefont {G.}~\bibnamefont {Juckeland}},
  \bibinfo {author} {\bibfnamefont {N.}~\bibnamefont {Biais}},\ and\ \bibinfo
  {author} {\bibfnamefont {V.}~\bibnamefont {Zaburdaev}},\ }\bibfield  {title}
  {\bibinfo {title} {Multiscale modeling of bacterial colonies: how pili
  mediate the dynamics of single cells and cellular aggregates},\ }\href
  {https://doi.org/10.1088/1367-2630/aa5483} {\bibfield  {journal} {\bibinfo
  {journal} {New J. Phys.}\ }\textbf {\bibinfo {volume} {19}},\ \bibinfo
  {pages} {015003} (\bibinfo {year} {2017})}\BibitemShut {NoStop}%
\bibitem [{\citenamefont {Fernández}\ \emph {et~al.}(2021)\citenamefont
  {Fernández} \emph {et~al.}}]{Fernandez2021}%
  \BibitemOpen
  \bibfield  {author} {\bibinfo {author} {\bibfnamefont {P.}~\bibnamefont
  {Fernández}} \emph {et~al.},\ }\bibfield  {title} {\bibinfo {title}
  {Surface-tension-induced budding drives alveologenesis in human mammary gland
  organoids},\ }\href {https://doi.org/10.1038/s41567-021-01336-7} {\bibfield
  {journal} {\bibinfo  {journal} {Nat. Phys.}\ }\textbf {\bibinfo {volume}
  {17}},\ \bibinfo {pages} {1130} (\bibinfo {year} {2021})}\BibitemShut
  {NoStop}%
\bibitem [{\citenamefont {Onesto}\ \emph {et~al.}(2024)\citenamefont {Onesto},
  \citenamefont {il~Kim},\ and\ \citenamefont {Pasca}}]{Pasca2024}%
  \BibitemOpen
  \bibfield  {author} {\bibinfo {author} {\bibfnamefont {M.~M.}\ \bibnamefont
  {Onesto}}, \bibinfo {author} {\bibfnamefont {J.}~\bibnamefont {il~Kim}},\
  and\ \bibinfo {author} {\bibfnamefont {S.~P.}\ \bibnamefont {Pasca}},\
  }\bibfield  {title} {\bibinfo {title} {Assembloid models of cell-cell
  interaction to study tissue and disease biology},\ }\href
  {https://doi.org/10.1016/j.stem.2024.09.017} {\bibfield  {journal} {\bibinfo
  {journal} {Cell Stem Cell}\ }\textbf {\bibinfo {volume} {31}},\ \bibinfo
  {pages} {1563} (\bibinfo {year} {2024})}\BibitemShut {NoStop}%
\bibitem [{\citenamefont {Wang}\ \emph {et~al.}(2025)\citenamefont {Wang} \emph
  {et~al.}}]{Wang2025}%
  \BibitemOpen
  \bibfield  {author} {\bibinfo {author} {\bibfnamefont {L.}~\bibnamefont
  {Wang}} \emph {et~al.},\ }\bibfield  {title} {\bibinfo {title} {Molecular and
  cellular dynamics of the developing human neocortex},\ }\bibfield  {journal}
  {\bibinfo  {journal} {Nature}\ }\href
  {https://doi.org/10.1038/s41586-024-08351-7} {10.1038/s41586-024-08351-7}
  (\bibinfo {year} {2025})\BibitemShut {NoStop}%
\bibitem [{\citenamefont {Sanchez}\ \emph {et~al.}(2012)\citenamefont {Sanchez}
  \emph {et~al.}}]{Sanchez2012}%
  \BibitemOpen
  \bibfield  {author} {\bibinfo {author} {\bibfnamefont {T.}~\bibnamefont
  {Sanchez}} \emph {et~al.},\ }\bibfield  {title} {\bibinfo {title}
  {Spontaneous motion in hierarchically assembled active matter},\ }\href
  {https://doi.org/10.1038/nature11591} {\bibfield  {journal} {\bibinfo
  {journal} {Nature}\ }\textbf {\bibinfo {volume} {491}},\ \bibinfo {pages}
  {431} (\bibinfo {year} {2012})}\BibitemShut {NoStop}%
\bibitem [{\citenamefont {Alert}\ and\ \citenamefont
  {Trepat}(2020)}]{Alert2020}%
  \BibitemOpen
  \bibfield  {author} {\bibinfo {author} {\bibfnamefont {R.}~\bibnamefont
  {Alert}}\ and\ \bibinfo {author} {\bibfnamefont {X.}~\bibnamefont {Trepat}},\
  }\bibfield  {title} {\bibinfo {title} {Physical models of collective cell
  migration},\ }\href {https://doi.org/10.1146/annurev-conmatphys-
  031218-013516} {\bibfield  {journal} {\bibinfo  {journal} {Annu. Rev.
  Condens. Matter Phys.}\ }\textbf {\bibinfo {volume} {11}},\ \bibinfo {pages}
  {77} (\bibinfo {year} {2020})}\BibitemShut {NoStop}%
\bibitem [{\citenamefont {Onsager}(1931)}]{Onsager}%
  \BibitemOpen
  \bibfield  {author} {\bibinfo {author} {\bibfnamefont {L.}~\bibnamefont
  {Onsager}},\ }\bibfield  {title} {\bibinfo {title} {Reciprocal relations in
  irreversible processes. ii.},\ }\href
  {https://doi.org/10.1103/PhysRev.38.2265} {\bibfield  {journal} {\bibinfo
  {journal} {Phys. Rev.}\ }\textbf {\bibinfo {volume} {38}},\ \bibinfo {pages}
  {2265} (\bibinfo {year} {1931})}\BibitemShut {NoStop}%
\bibitem [{\citenamefont {Pathria}\ and\ \citenamefont
  {Beale}(2011)}]{Pathria}%
  \BibitemOpen
  \bibfield  {author} {\bibinfo {author} {\bibfnamefont {R.~K.}\ \bibnamefont
  {Pathria}}\ and\ \bibinfo {author} {\bibfnamefont {P.~D.}\ \bibnamefont
  {Beale}},\ }\href@noop {} {\emph {\bibinfo {title} {Statistical
  Mechanics}}},\ \bibinfo {edition} {3rd}\ ed.\ (\bibinfo  {publisher}
  {Elsevier},\ \bibinfo {address} {Amsterdam},\ \bibinfo {year}
  {2011})\BibitemShut {NoStop}%
\bibitem [{\citenamefont {Bertini}\ \emph {et~al.}(2015)\citenamefont
  {Bertini}, \citenamefont {De~Sole}, \citenamefont {Gabrielli}, \citenamefont
  {Jona-Lasinio},\ and\ \citenamefont {Landim}}]{Bertini_RMP2015}%
  \BibitemOpen
  \bibfield  {author} {\bibinfo {author} {\bibfnamefont {L.}~\bibnamefont
  {Bertini}}, \bibinfo {author} {\bibfnamefont {A.}~\bibnamefont {De~Sole}},
  \bibinfo {author} {\bibfnamefont {D.}~\bibnamefont {Gabrielli}}, \bibinfo
  {author} {\bibfnamefont {G.}~\bibnamefont {Jona-Lasinio}},\ and\ \bibinfo
  {author} {\bibfnamefont {C.}~\bibnamefont {Landim}},\ }\bibfield  {title}
  {\bibinfo {title} {Macroscopic fluctuation theory},\ }\href
  {https://doi.org/10.1103/RevModPhys.87.593} {\bibfield  {journal} {\bibinfo
  {journal} {Rev. Mod. Phys.}\ }\textbf {\bibinfo {volume} {87}},\ \bibinfo
  {pages} {593} (\bibinfo {year} {2015})}\BibitemShut {NoStop}%
\bibitem [{\citenamefont {Gardiner}(1990)}]{Gardinerbook}%
  \BibitemOpen
  \bibfield  {author} {\bibinfo {author} {\bibfnamefont {C.~W.}\ \bibnamefont
  {Gardiner}},\ }\href@noop {} {\emph {\bibinfo {title} {Handbook of Stochastic
  Methods: For Physics, Chemistry and the Natural Sciences}}}\ (\bibinfo
  {publisher} {Springer Berlin Heidelberg},\ \bibinfo {year}
  {1990})\BibitemShut {NoStop}%
\bibitem [{\citenamefont {Gnesotto1}\ \emph {et~al.}(2018)\citenamefont
  {Gnesotto1}, \citenamefont {Mura1}, \citenamefont {Gladrow},\ and\
  \citenamefont {Broedersz}}]{Broedersz2018}%
  \BibitemOpen
  \bibfield  {author} {\bibinfo {author} {\bibfnamefont {F.~S.}\ \bibnamefont
  {Gnesotto1}}, \bibinfo {author} {\bibfnamefont {F.}~\bibnamefont {Mura1}},
  \bibinfo {author} {\bibfnamefont {J.}~\bibnamefont {Gladrow}},\ and\ \bibinfo
  {author} {\bibfnamefont {C.~P.}\ \bibnamefont {Broedersz}},\ }\bibfield
  {title} {\bibinfo {title} {Broken detailed balance and non-equilibrium
  dynamics in living systems: a review},\ }\href
  {https://doi.org/10.1088/1361-6633/aab3ed} {\bibfield  {journal} {\bibinfo
  {journal} {Rep. Prog. Phys.}\ }\textbf {\bibinfo {volume} {81}},\ \bibinfo
  {pages} {066601} (\bibinfo {year} {2018})}\BibitemShut {NoStop}%
\bibitem [{\citenamefont {Thapa}\ \emph {et~al.}(2024)\citenamefont {Thapa}
  \emph {et~al.}}]{Broedersz2024}%
  \BibitemOpen
  \bibfield  {author} {\bibinfo {author} {\bibfnamefont {S.}~\bibnamefont
  {Thapa}} \emph {et~al.},\ }\bibfield  {title} {\bibinfo {title}
  {Nonequilibrium probability currents in optically-driven colloidal
  suspensions},\ }\href {https://doi.org/10.21468/SciPostPhys.17.4.096}
  {\bibfield  {journal} {\bibinfo  {journal} {SciPost Phys.}\ }\textbf
  {\bibinfo {volume} {17}},\ \bibinfo {pages} {096} (\bibinfo {year}
  {2024})}\BibitemShut {NoStop}%
\bibitem [{\citenamefont {Ascione}\ \emph {et~al.}(2024)\citenamefont {Ascione}
  \emph {et~al.}}]{Ascione2024}%
  \BibitemOpen
  \bibfield  {author} {\bibinfo {author} {\bibfnamefont {F.}~\bibnamefont
  {Ascione}} \emph {et~al.},\ }\bibfield  {title} {\bibinfo {title}
  {Gradient-induced instability in tumour spheroids unveils the impact of
  microenvironmental nutrient changes},\ }\href
  {https://doi.org/10.1038/s41598-024-69570-6} {\bibfield  {journal} {\bibinfo
  {journal} {Sci. Rep.}\ }\textbf {\bibinfo {volume} {14}},\ \bibinfo {pages}
  {20837} (\bibinfo {year} {2024})}\BibitemShut {NoStop}%
\bibitem [{\citenamefont {Xu}\ \emph {et~al.}(2019)\citenamefont {Xu} \emph
  {et~al.}}]{Xu2019}%
  \BibitemOpen
  \bibfield  {author} {\bibinfo {author} {\bibfnamefont {H.}~\bibnamefont {Xu}}
  \emph {et~al.},\ }\bibfield  {title} {\bibinfo {title} {Self-organization of
  swimmers drives long-range fluid transport in bacterial colonies},\ }\href
  {https://doi.org/10.1038/s41467-019-09818-2} {\bibfield  {journal} {\bibinfo
  {journal} {Nat. Commun.}\ }\textbf {\bibinfo {volume} {10}},\ \bibinfo
  {pages} {1792} (\bibinfo {year} {2019})}\BibitemShut {NoStop}%
\bibitem [{\citenamefont {P{\"o}nisch}\ \emph {et~al.}(2018)\citenamefont
  {P{\"o}nisch}, \citenamefont {Eckenrode}, \citenamefont {Alzurqa},
  \citenamefont {Nasrollahi}, \citenamefont {Weber}, \citenamefont
  {Zaburdaev},\ and\ \citenamefont {Biais}}]{ponisch2018pili}%
  \BibitemOpen
  \bibfield  {author} {\bibinfo {author} {\bibfnamefont {W.}~\bibnamefont
  {P{\"o}nisch}}, \bibinfo {author} {\bibfnamefont {K.~B.}\ \bibnamefont
  {Eckenrode}}, \bibinfo {author} {\bibfnamefont {K.}~\bibnamefont {Alzurqa}},
  \bibinfo {author} {\bibfnamefont {H.}~\bibnamefont {Nasrollahi}}, \bibinfo
  {author} {\bibfnamefont {C.}~\bibnamefont {Weber}}, \bibinfo {author}
  {\bibfnamefont {V.}~\bibnamefont {Zaburdaev}},\ and\ \bibinfo {author}
  {\bibfnamefont {N.}~\bibnamefont {Biais}},\ }\bibfield  {title} {\bibinfo
  {title} {Pili mediated intercellular forces shape heterogeneous bacterial
  microcolonies prior to multicellular differentiation},\ }\href
  {https://doi.org/10.1038/s41598-018-34754-4} {\bibfield  {journal} {\bibinfo
  {journal} {Sci. Rep.}\ }\textbf {\bibinfo {volume} {8}},\ \bibinfo {pages}
  {16567} (\bibinfo {year} {2018})}\BibitemShut {NoStop}%
\bibitem [{\citenamefont {Tortorella}\ \emph {et~al.}(2022)\citenamefont
  {Tortorella}, \citenamefont {Argentati}, \citenamefont {Emiliani},
  \citenamefont {Martino},\ and\ \citenamefont {Morena}}]{tortorella2021role}%
  \BibitemOpen
  \bibfield  {author} {\bibinfo {author} {\bibfnamefont {I.}~\bibnamefont
  {Tortorella}}, \bibinfo {author} {\bibfnamefont {C.}~\bibnamefont
  {Argentati}}, \bibinfo {author} {\bibfnamefont {C.}~\bibnamefont {Emiliani}},
  \bibinfo {author} {\bibfnamefont {S.}~\bibnamefont {Martino}},\ and\ \bibinfo
  {author} {\bibfnamefont {F.}~\bibnamefont {Morena}},\ }\bibfield  {title}
  {\bibinfo {title} {The role of physical cues in the development of stem
  cell-derived organoids},\ }\href {https://doi.org/10.1007/s00249-021-01551-3}
  {\bibfield  {journal} {\bibinfo  {journal} {Eur. Biophys. J.}\ }\textbf
  {\bibinfo {volume} {51}},\ \bibinfo {pages} {105–117} (\bibinfo {year}
  {2022})}\BibitemShut {NoStop}%
\bibitem [{\citenamefont {Cronenberg~T}(2021)}]{Maier2021}%
  \BibitemOpen
  \bibfield  {author} {\bibinfo {author} {\bibfnamefont {W.~I. M.~B.}\
  \bibnamefont {Cronenberg~T}, \bibfnamefont {Hennes~M}},\ }\bibfield  {title}
  {\bibinfo {title} {Antibiotics modulate attractive interactions in bacterial
  colonies affecting survivability under combined treatment},\ }\href
  {https://doi.org/10.1371/ journal.ppat.1009251} {\bibfield  {journal}
  {\bibinfo  {journal} {PLoS Pathog.}\ }\textbf {\bibinfo {volume} {17}},\
  \bibinfo {pages} {e1009251} (\bibinfo {year} {2021})}\BibitemShut {NoStop}%
\bibitem [{\citenamefont {Craig}\ \emph {et~al.}(2004)\citenamefont {Craig},
  \citenamefont {Pique},\ and\ \citenamefont {Tainer}}]{Tainer2004}%
  \BibitemOpen
  \bibfield  {author} {\bibinfo {author} {\bibfnamefont {L.}~\bibnamefont
  {Craig}}, \bibinfo {author} {\bibfnamefont {M.~E.}\ \bibnamefont {Pique}},\
  and\ \bibinfo {author} {\bibfnamefont {J.~A.}\ \bibnamefont {Tainer}},\
  }\bibfield  {title} {\bibinfo {title} {Type iv pilus structure and bacterial
  pathogenicity},\ }\href {https://doi.org/10.1038/nrmicro885} {\bibfield
  {journal} {\bibinfo  {journal} {Nat. Rev. Microbiol.}\ }\textbf {\bibinfo
  {volume} {2}},\ \bibinfo {pages} {363} (\bibinfo {year} {2004})}\BibitemShut
  {NoStop}%
\bibitem [{\citenamefont {Mattick}(2002)}]{Mattick2002}%
  \BibitemOpen
  \bibfield  {author} {\bibinfo {author} {\bibfnamefont {J.~S.}\ \bibnamefont
  {Mattick}},\ }\bibfield  {title} {\bibinfo {title} {Type iv pili and
  twitching motility},\ }\href
  {https://doi.org/10.1146/annurev.micro.56.012302.160938} {\bibfield
  {journal} {\bibinfo  {journal} {Annu. Rev. Microbiol.}\ }\textbf {\bibinfo
  {volume} {56}},\ \bibinfo {pages} {289} (\bibinfo {year} {2002})}\BibitemShut
  {NoStop}%
\bibitem [{\citenamefont {Holz}\ \emph {et~al.}(2010)\citenamefont {Holz},
  \citenamefont {Opitz}, \citenamefont {Greune}, \citenamefont {Kurre},
  \citenamefont {Koomey}, \citenamefont {Schmidt},\ and\ \citenamefont
  {Maier}}]{Maier_PRL2010}%
  \BibitemOpen
  \bibfield  {author} {\bibinfo {author} {\bibfnamefont {C.}~\bibnamefont
  {Holz}}, \bibinfo {author} {\bibfnamefont {D.}~\bibnamefont {Opitz}},
  \bibinfo {author} {\bibfnamefont {L.}~\bibnamefont {Greune}}, \bibinfo
  {author} {\bibfnamefont {R.}~\bibnamefont {Kurre}}, \bibinfo {author}
  {\bibfnamefont {M.}~\bibnamefont {Koomey}}, \bibinfo {author} {\bibfnamefont
  {M.~A.}\ \bibnamefont {Schmidt}},\ and\ \bibinfo {author} {\bibfnamefont
  {B.}~\bibnamefont {Maier}},\ }\bibfield  {title} {\bibinfo {title} {Multiple
  pilus motors cooperate for persistent bacterial movement in two dimensions},\
  }\href {https://doi.org/10.1103/PhysRevLett.104.178104} {\bibfield  {journal}
  {\bibinfo  {journal} {Phys. Rev. Lett.}\ }\textbf {\bibinfo {volume} {104}},\
  \bibinfo {pages} {178104} (\bibinfo {year} {2010})}\BibitemShut {NoStop}%
\bibitem [{\citenamefont {Merz}\ \emph {et~al.}(2000)\citenamefont {Merz},
  \citenamefont {So},\ and\ \citenamefont {Sheetz}}]{SheetzMP_Nature2000}%
  \BibitemOpen
  \bibfield  {author} {\bibinfo {author} {\bibfnamefont {A.}~\bibnamefont
  {Merz}}, \bibinfo {author} {\bibfnamefont {M.}~\bibnamefont {So}},\ and\
  \bibinfo {author} {\bibfnamefont {M.}~\bibnamefont {Sheetz}},\ }\bibfield
  {title} {\bibinfo {title} {Pilus retraction powers bacterial twitching
  motility},\ }\href {https://doi.org/10.1038/35024105} {\bibfield  {journal}
  {\bibinfo  {journal} {Nature}\ }\textbf {\bibinfo {volume} {407}},\ \bibinfo
  {pages} {98} (\bibinfo {year} {2000})}\BibitemShut {NoStop}%
\bibitem [{\citenamefont {Skerker}\ and\ \citenamefont
  {Berg}(2001)}]{Berg_PNAS2001}%
  \BibitemOpen
  \bibfield  {author} {\bibinfo {author} {\bibfnamefont {J.~M.}\ \bibnamefont
  {Skerker}}\ and\ \bibinfo {author} {\bibfnamefont {H.~C.}\ \bibnamefont
  {Berg}},\ }\bibfield  {title} {\bibinfo {title} {Direct observation of
  extension and retraction of type iv pili},\ }\href
  {https://doi.org/10.1073/pnas.121171698} {\bibfield  {journal} {\bibinfo
  {journal} {Proc. Natl. Acad. Sci. U.S.A.}\ }\textbf {\bibinfo {volume}
  {98}},\ \bibinfo {pages} {6901} (\bibinfo {year} {2001})}\BibitemShut
  {NoStop}%
\bibitem [{\citenamefont {Craig}\ \emph {et~al.}(2019)\citenamefont {Craig},
  \citenamefont {Forest},\ and\ \citenamefont {Maier}}]{Maier2019}%
  \BibitemOpen
  \bibfield  {author} {\bibinfo {author} {\bibfnamefont {L.}~\bibnamefont
  {Craig}}, \bibinfo {author} {\bibfnamefont {K.~T.}\ \bibnamefont {Forest}},\
  and\ \bibinfo {author} {\bibfnamefont {B.}~\bibnamefont {Maier}},\ }\bibfield
   {title} {\bibinfo {title} {Type iv pili: dynamics, biophysics and functional
  consequences},\ }\href {https://doi.org/10.1038/s41579-019-0195-4} {\bibfield
   {journal} {\bibinfo  {journal} {Nat. Rev. Microbiol.}\ }\textbf {\bibinfo
  {volume} {17}},\ \bibinfo {pages} {429} (\bibinfo {year} {2019})}\BibitemShut
  {NoStop}%
\bibitem [{\citenamefont {Bertini}\ \emph {et~al.}(2001)\citenamefont
  {Bertini}, \citenamefont {De~Sole}, \citenamefont {Gabrielli}, \citenamefont
  {Jona-Lasinio},\ and\ \citenamefont {Landim}}]{Bertini_PRL2001}%
  \BibitemOpen
  \bibfield  {author} {\bibinfo {author} {\bibfnamefont {L.}~\bibnamefont
  {Bertini}}, \bibinfo {author} {\bibfnamefont {A.}~\bibnamefont {De~Sole}},
  \bibinfo {author} {\bibfnamefont {D.}~\bibnamefont {Gabrielli}}, \bibinfo
  {author} {\bibfnamefont {G.}~\bibnamefont {Jona-Lasinio}},\ and\ \bibinfo
  {author} {\bibfnamefont {C.}~\bibnamefont {Landim}},\ }\bibfield  {title}
  {\bibinfo {title} {Fluctuations in stationary nonequilibrium states of
  irreversible processes},\ }\href
  {https://doi.org/10.1103/PhysRevLett.87.040601} {\bibfield  {journal}
  {\bibinfo  {journal} {Phys. Rev. Lett.}\ }\textbf {\bibinfo {volume} {87}},\
  \bibinfo {pages} {040601} (\bibinfo {year} {2001})}\BibitemShut {NoStop}%
\bibitem [{\citenamefont {Derrida}(2007)}]{Derrida_JSM2007}%
  \BibitemOpen
  \bibfield  {author} {\bibinfo {author} {\bibfnamefont {B.}~\bibnamefont
  {Derrida}},\ }\bibfield  {title} {\bibinfo {title} {Non-equilibrium steady
  states: fluctuations and large deviations of the density and of the
  current},\ }\href {https://doi.org/10.1088/1742-5468/2007/07/p07023}
  {\bibfield  {journal} {\bibinfo  {journal} {J. Stat. Mech.}\ }\textbf
  {\bibinfo {volume} {2007}},\ \bibinfo {pages} {P07023} (\bibinfo {year}
  {2007})}\BibitemShut {NoStop}%
\bibitem [{\citenamefont {Chakraborti}\ and\ \citenamefont
  {Zaburdaev}(2024)}]{Chakraborti-PRR2024}%
  \BibitemOpen
  \bibfield  {author} {\bibinfo {author} {\bibfnamefont {S.}~\bibnamefont
  {Chakraborti}}\ and\ \bibinfo {author} {\bibfnamefont {V.}~\bibnamefont
  {Zaburdaev}},\ }\bibfield  {title} {\bibinfo {title} {Transport in cellular
  aggregates described by fluctuating hydrodynamics},\ }\href
  {https://doi.org/10.1103/PhysRevResearch.6.043064} {\bibfield  {journal}
  {\bibinfo  {journal} {Phys. Rev. Res.}\ }\textbf {\bibinfo {volume} {6}},\
  \bibinfo {pages} {043064} (\bibinfo {year} {2024})}\BibitemShut {NoStop}%
\bibitem [{\citenamefont {Cates}\ and\ \citenamefont
  {Tailleur}(2015)}]{Cates_review}%
  \BibitemOpen
  \bibfield  {author} {\bibinfo {author} {\bibfnamefont {M.~E.}\ \bibnamefont
  {Cates}}\ and\ \bibinfo {author} {\bibfnamefont {J.}~\bibnamefont
  {Tailleur}},\ }\bibfield  {title} {\bibinfo {title} {Motility-induced phase
  separation},\ }\href
  {https://doi.org/https://doi.org/10.1146/annurev-conmatphys-031214-014710}
  {\bibfield  {journal} {\bibinfo  {journal} {Annu. Rev. Condens. Matter
  Phys.}\ }\textbf {\bibinfo {volume} {6}},\ \bibinfo {pages} {219} (\bibinfo
  {year} {2015})}\BibitemShut {NoStop}%
\bibitem [{\citenamefont {Ellison}\ \emph {et~al.}(2021)\citenamefont
  {Ellison}, \citenamefont {Whitfield},\ and\ \citenamefont {Brun}}]{Ellison}%
  \BibitemOpen
  \bibfield  {author} {\bibinfo {author} {\bibfnamefont {C.~K.}\ \bibnamefont
  {Ellison}}, \bibinfo {author} {\bibfnamefont {G.~B.}\ \bibnamefont
  {Whitfield}},\ and\ \bibinfo {author} {\bibfnamefont {Y.~V.}\ \bibnamefont
  {Brun}},\ }\bibfield  {title} {\bibinfo {title} {{Type IV Pili: dynamic
  bacterial nanomachines}},\ }\href {https://doi.org/10.1093/femsre/fuab053}
  {\bibfield  {journal} {\bibinfo  {journal} {FEMS Microbiology Reviews}\
  }\textbf {\bibinfo {volume} {46}},\ \bibinfo {pages} {fuab053} (\bibinfo
  {year} {2021})}\BibitemShut {NoStop}%
\bibitem [{\citenamefont {Henrichsen}(1983)}]{Henrichsen}%
  \BibitemOpen
  \bibfield  {author} {\bibinfo {author} {\bibfnamefont {J.}~\bibnamefont
  {Henrichsen}},\ }\bibfield  {title} {\bibinfo {title} {Twitching motility},\
  }\href {https://doi.org/10.1146/annurev.mi.37.100183.000501} {\bibfield
  {journal} {\bibinfo  {journal} {Annual Review of Microbiology}\ }\textbf
  {\bibinfo {volume} {37}},\ \bibinfo {pages} {81} (\bibinfo {year} {1983})},\
  \bibinfo {note} {pMID: 6139059}\BibitemShut {NoStop}%
\bibitem [{\citenamefont {Evans}\ and\ \citenamefont
  {Hanney}(2005)}]{Evans_JPA2005}%
  \BibitemOpen
  \bibfield  {author} {\bibinfo {author} {\bibfnamefont {M.~R.}\ \bibnamefont
  {Evans}}\ and\ \bibinfo {author} {\bibfnamefont {T.}~\bibnamefont {Hanney}},\
  }\bibfield  {title} {\bibinfo {title} {Nonequilibrium statistical mechanics
  of the zero-range process and related models},\ }\href
  {https://doi.org/10.1088/0305-4470/38/19/R01} {\bibfield  {journal} {\bibinfo
   {journal} {J. Phys. A: Math. Gen.}\ }\textbf {\bibinfo {volume} {38}},\
  \bibinfo {pages} {R195} (\bibinfo {year} {2005})}\BibitemShut {NoStop}%
\bibitem [{\citenamefont {Weber}\ \emph {et~al.}(2015)\citenamefont {Weber},
  \citenamefont {Lin}, \citenamefont {Biais},\ and\ \citenamefont
  {Zaburdaev}}]{Zaburdaev_PRE2015}%
  \BibitemOpen
  \bibfield  {author} {\bibinfo {author} {\bibfnamefont {C.~A.}\ \bibnamefont
  {Weber}}, \bibinfo {author} {\bibfnamefont {Y.~T.}\ \bibnamefont {Lin}},
  \bibinfo {author} {\bibfnamefont {N.}~\bibnamefont {Biais}},\ and\ \bibinfo
  {author} {\bibfnamefont {V.}~\bibnamefont {Zaburdaev}},\ }\bibfield  {title}
  {\bibinfo {title} {Formation and dissolution of bacterial colonies},\ }\href
  {https://doi.org/10.1103/PhysRevE.92.032704} {\bibfield  {journal} {\bibinfo
  {journal} {Phys. Rev. E}\ }\textbf {\bibinfo {volume} {92}},\ \bibinfo
  {pages} {032704} (\bibinfo {year} {2015})}\BibitemShut {NoStop}%
\bibitem [{\citenamefont {Kuan}\ \emph {et~al.}(2021)\citenamefont {Kuan},
  \citenamefont {P\"onisch}, \citenamefont {J\"ulicher},\ and\ \citenamefont
  {Zaburdaev}}]{Zaburdaev_PRL2021}%
  \BibitemOpen
  \bibfield  {author} {\bibinfo {author} {\bibfnamefont {H.-S.}\ \bibnamefont
  {Kuan}}, \bibinfo {author} {\bibfnamefont {W.}~\bibnamefont {P\"onisch}},
  \bibinfo {author} {\bibfnamefont {F.}~\bibnamefont {J\"ulicher}},\ and\
  \bibinfo {author} {\bibfnamefont {V.}~\bibnamefont {Zaburdaev}},\ }\bibfield
  {title} {\bibinfo {title} {Continuum theory of active phase separation in
  cellular aggregates},\ }\href
  {https://doi.org/10.1103/PhysRevLett.126.018102} {\bibfield  {journal}
  {\bibinfo  {journal} {Phys. Rev. Lett.}\ }\textbf {\bibinfo {volume} {126}},\
  \bibinfo {pages} {018102} (\bibinfo {year} {2021})}\BibitemShut {NoStop}%
\bibitem [{\citenamefont {Cates}\ and\ \citenamefont
  {Tailleur}(2013)}]{Cates_2013}%
  \BibitemOpen
  \bibfield  {author} {\bibinfo {author} {\bibfnamefont {M.~E.}\ \bibnamefont
  {Cates}}\ and\ \bibinfo {author} {\bibfnamefont {J.}~\bibnamefont
  {Tailleur}},\ }\bibfield  {title} {\bibinfo {title} {When are active brownian
  particles and run-and-tumble particles equivalent? consequences for
  motility-induced phase separation},\ }\href
  {https://doi.org/10.1209/0295-5075/101/20010} {\bibfield  {journal} {\bibinfo
   {journal} {EPL}\ }\textbf {\bibinfo {volume} {101}},\ \bibinfo {pages}
  {20010} (\bibinfo {year} {2013})}\BibitemShut {NoStop}%
\bibitem [{\citenamefont {Caprini}\ \emph {et~al.}(2020)\citenamefont
  {Caprini}, \citenamefont {Marini Bettolo~Marconi},\ and\ \citenamefont
  {Puglisi}}]{Caprini_2020}%
  \BibitemOpen
  \bibfield  {author} {\bibinfo {author} {\bibfnamefont {L.}~\bibnamefont
  {Caprini}}, \bibinfo {author} {\bibfnamefont {U.}~\bibnamefont {Marini
  Bettolo~Marconi}},\ and\ \bibinfo {author} {\bibfnamefont {A.}~\bibnamefont
  {Puglisi}},\ }\bibfield  {title} {\bibinfo {title} {Spontaneous velocity
  alignment in motility-induced phase separation},\ }\href
  {https://doi.org/10.1103/PhysRevLett.124.078001} {\bibfield  {journal}
  {\bibinfo  {journal} {Phys. Rev. Lett.}\ }\textbf {\bibinfo {volume} {124}},\
  \bibinfo {pages} {078001} (\bibinfo {year} {2020})}\BibitemShut {NoStop}%
\bibitem [{\citenamefont {Caporusso}\ \emph {et~al.}(2020)\citenamefont
  {Caporusso}, \citenamefont {Digregorio}, \citenamefont {Levis}, \citenamefont
  {Cugliandolo},\ and\ \citenamefont {Gonnella}}]{Caporusso_2020}%
  \BibitemOpen
  \bibfield  {author} {\bibinfo {author} {\bibfnamefont {C.~B.}\ \bibnamefont
  {Caporusso}}, \bibinfo {author} {\bibfnamefont {P.}~\bibnamefont
  {Digregorio}}, \bibinfo {author} {\bibfnamefont {D.}~\bibnamefont {Levis}},
  \bibinfo {author} {\bibfnamefont {L.~F.}\ \bibnamefont {Cugliandolo}},\ and\
  \bibinfo {author} {\bibfnamefont {G.}~\bibnamefont {Gonnella}},\ }\bibfield
  {title} {\bibinfo {title} {Motility-induced microphase and macrophase
  separation in a two-dimensional active brownian particle system},\ }\href
  {https://doi.org/10.1103/PhysRevLett.125.178004} {\bibfield  {journal}
  {\bibinfo  {journal} {Phys. Rev. Lett.}\ }\textbf {\bibinfo {volume} {125}},\
  \bibinfo {pages} {178004} (\bibinfo {year} {2020})}\BibitemShut {NoStop}%
\bibitem [{\citenamefont {Dandekar}\ \emph {et~al.}(2020)\citenamefont
  {Dandekar}, \citenamefont {Chakraborti},\ and\ \citenamefont
  {Rajesh}}]{Rahul_PRE2020}%
  \BibitemOpen
  \bibfield  {author} {\bibinfo {author} {\bibfnamefont {R.}~\bibnamefont
  {Dandekar}}, \bibinfo {author} {\bibfnamefont {S.}~\bibnamefont
  {Chakraborti}},\ and\ \bibinfo {author} {\bibfnamefont {R.}~\bibnamefont
  {Rajesh}},\ }\bibfield  {title} {\bibinfo {title} {Hard core run and tumble
  particles on a one-dimensional lattice},\ }\href
  {https://doi.org/10.1103/PhysRevE.102.062111} {\bibfield  {journal} {\bibinfo
   {journal} {Phys. Rev. E}\ }\textbf {\bibinfo {volume} {102}},\ \bibinfo
  {pages} {062111} (\bibinfo {year} {2020})}\BibitemShut {NoStop}%
\bibitem [{\citenamefont {Redner}\ \emph {et~al.}(2013)\citenamefont {Redner},
  \citenamefont {Baskaran},\ and\ \citenamefont {Hagan}}]{Redner_2013}%
  \BibitemOpen
  \bibfield  {author} {\bibinfo {author} {\bibfnamefont {G.~S.}\ \bibnamefont
  {Redner}}, \bibinfo {author} {\bibfnamefont {A.}~\bibnamefont {Baskaran}},\
  and\ \bibinfo {author} {\bibfnamefont {M.~F.}\ \bibnamefont {Hagan}},\
  }\bibfield  {title} {\bibinfo {title} {Reentrant phase behavior in active
  colloids with attraction},\ }\href
  {https://doi.org/10.1103/PhysRevE.88.012305} {\bibfield  {journal} {\bibinfo
  {journal} {Phys. Rev. E}\ }\textbf {\bibinfo {volume} {88}},\ \bibinfo
  {pages} {012305} (\bibinfo {year} {2013})}\BibitemShut {NoStop}%
\bibitem [{\citenamefont {Caprini}\ and\ \citenamefont
  {L\"owen}(2023)}]{Caprini_2023}%
  \BibitemOpen
  \bibfield  {author} {\bibinfo {author} {\bibfnamefont {L.}~\bibnamefont
  {Caprini}}\ and\ \bibinfo {author} {\bibfnamefont {H.}~\bibnamefont
  {L\"owen}},\ }\bibfield  {title} {\bibinfo {title} {Flocking without
  alignment interactions in attractive active brownian particles},\ }\href
  {https://doi.org/10.1103/PhysRevLett.130.148202} {\bibfield  {journal}
  {\bibinfo  {journal} {Phys. Rev. Lett.}\ }\textbf {\bibinfo {volume} {130}},\
  \bibinfo {pages} {148202} (\bibinfo {year} {2023})}\BibitemShut {NoStop}%
\bibitem [{\citenamefont {Alston}\ \emph {et~al.}(2022)\citenamefont {Alston},
  \citenamefont {Parry}, \citenamefont {Voituriez},\ and\ \citenamefont
  {Bertrand}}]{Rafael_PRE2022}%
  \BibitemOpen
  \bibfield  {author} {\bibinfo {author} {\bibfnamefont {H.}~\bibnamefont
  {Alston}}, \bibinfo {author} {\bibfnamefont {A.~O.}\ \bibnamefont {Parry}},
  \bibinfo {author} {\bibfnamefont {R.}~\bibnamefont {Voituriez}},\ and\
  \bibinfo {author} {\bibfnamefont {T.}~\bibnamefont {Bertrand}},\ }\bibfield
  {title} {\bibinfo {title} {Intermittent attractive interactions lead to
  microphase separation in nonmotile active matter},\ }\href
  {https://doi.org/10.1103/PhysRevE.106.034603} {\bibfield  {journal} {\bibinfo
   {journal} {Phys. Rev. E}\ }\textbf {\bibinfo {volume} {106}},\ \bibinfo
  {pages} {034603} (\bibinfo {year} {2022})}\BibitemShut {NoStop}%
\bibitem [{\citenamefont {Majumdar}\ \emph {et~al.}(1998)\citenamefont
  {Majumdar}, \citenamefont {Krishnamurthy},\ and\ \citenamefont
  {Barma}}]{Barma-model}%
  \BibitemOpen
  \bibfield  {author} {\bibinfo {author} {\bibfnamefont {S.~N.}\ \bibnamefont
  {Majumdar}}, \bibinfo {author} {\bibfnamefont {S.}~\bibnamefont
  {Krishnamurthy}},\ and\ \bibinfo {author} {\bibfnamefont {M.}~\bibnamefont
  {Barma}},\ }\bibfield  {title} {\bibinfo {title} {Nonequilibrium phase
  transitions in models of aggregation, adsorption, and dissociation},\ }\href
  {https://doi.org/10.1103/PhysRevLett.81.3691} {\bibfield  {journal} {\bibinfo
   {journal} {Phys. Rev. Lett.}\ }\textbf {\bibinfo {volume} {81}},\ \bibinfo
  {pages} {3691} (\bibinfo {year} {1998})}\BibitemShut {NoStop}%
\bibitem [{\citenamefont {Chakraborti}\ \emph {et~al.}(2021)\citenamefont
  {Chakraborti}, \citenamefont {Chakraborty}, \citenamefont {Das},
  \citenamefont {Dandekar},\ and\ \citenamefont
  {Pradhan}}]{Chakraborti_PRE2021}%
  \BibitemOpen
  \bibfield  {author} {\bibinfo {author} {\bibfnamefont {S.}~\bibnamefont
  {Chakraborti}}, \bibinfo {author} {\bibfnamefont {T.}~\bibnamefont
  {Chakraborty}}, \bibinfo {author} {\bibfnamefont {A.}~\bibnamefont {Das}},
  \bibinfo {author} {\bibfnamefont {R.}~\bibnamefont {Dandekar}},\ and\
  \bibinfo {author} {\bibfnamefont {P.}~\bibnamefont {Pradhan}},\ }\bibfield
  {title} {\bibinfo {title} {Transport and fluctuations in mass aggregation
  processes: Mobility-driven clustering},\ }\href
  {https://doi.org/10.1103/PhysRevE.103.042133} {\bibfield  {journal} {\bibinfo
   {journal} {Phys. Rev. E}\ }\textbf {\bibinfo {volume} {103}},\ \bibinfo
  {pages} {042133} (\bibinfo {year} {2021})}\BibitemShut {NoStop}%
\bibitem [{\citenamefont {Bodineau}\ and\ \citenamefont
  {Derrida}(2004)}]{Derrida_PRL2004}%
  \BibitemOpen
  \bibfield  {author} {\bibinfo {author} {\bibfnamefont {T.}~\bibnamefont
  {Bodineau}}\ and\ \bibinfo {author} {\bibfnamefont {B.}~\bibnamefont
  {Derrida}},\ }\bibfield  {title} {\bibinfo {title} {Current fluctuations in
  nonequilibrium diffusive systems: An additivity principle},\ }\href
  {https://doi.org/10.1103/PhysRevLett.92.180601} {\bibfield  {journal}
  {\bibinfo  {journal} {Phys. Rev. Lett.}\ }\textbf {\bibinfo {volume} {92}},\
  \bibinfo {pages} {180601} (\bibinfo {year} {2004})}\BibitemShut {NoStop}%
\bibitem [{\citenamefont {Arita}\ \emph {et~al.}(2014)\citenamefont {Arita},
  \citenamefont {Krapivsky},\ and\ \citenamefont
  {Mallick}}]{Krapivsky_PRE2014}%
  \BibitemOpen
  \bibfield  {author} {\bibinfo {author} {\bibfnamefont {C.}~\bibnamefont
  {Arita}}, \bibinfo {author} {\bibfnamefont {P.~L.}\ \bibnamefont
  {Krapivsky}},\ and\ \bibinfo {author} {\bibfnamefont {K.}~\bibnamefont
  {Mallick}},\ }\bibfield  {title} {\bibinfo {title} {Generalized exclusion
  processes: Transport coefficients},\ }\href
  {https://doi.org/10.1103/PhysRevE.90.052108} {\bibfield  {journal} {\bibinfo
  {journal} {Phys. Rev. E}\ }\textbf {\bibinfo {volume} {90}},\ \bibinfo
  {pages} {052108} (\bibinfo {year} {2014})}\BibitemShut {NoStop}%
\bibitem [{sup()}]{supplement}%
  \BibitemOpen
  \bibfield  {title} {\bibinfo {title} {Please see the appendices for detail},\
  }\href@noop {} {\ }\BibitemShut {NoStop}%
\bibitem [{\citenamefont {Alexander}\ and\ \citenamefont
  {Pincus}(1978)}]{Alexander_PRB1978}%
  \BibitemOpen
  \bibfield  {author} {\bibinfo {author} {\bibfnamefont {S.}~\bibnamefont
  {Alexander}}\ and\ \bibinfo {author} {\bibfnamefont {P.}~\bibnamefont
  {Pincus}},\ }\bibfield  {title} {\bibinfo {title} {Diffusion of labeled
  particles on one-dimensional chains},\ }\href
  {https://doi.org/10.1103/PhysRevB.18.2011} {\bibfield  {journal} {\bibinfo
  {journal} {Phys. Rev. B}\ }\textbf {\bibinfo {volume} {18}},\ \bibinfo
  {pages} {2011} (\bibinfo {year} {1978})}\BibitemShut {NoStop}%
\bibitem [{\citenamefont {Krapivsky}\ \emph {et~al.}(2015)\citenamefont
  {Krapivsky}, \citenamefont {Mallick},\ and\ \citenamefont
  {Sadhu}}]{Krapivsky_JSP2015}%
  \BibitemOpen
  \bibfield  {author} {\bibinfo {author} {\bibfnamefont {P.~L.}\ \bibnamefont
  {Krapivsky}}, \bibinfo {author} {\bibfnamefont {K.}~\bibnamefont {Mallick}},\
  and\ \bibinfo {author} {\bibfnamefont {T.}~\bibnamefont {Sadhu}},\ }\bibfield
   {title} {\bibinfo {title} {Tagged particle in single-file diffusion},\
  }\href {https://doi.org/10.1007/s10955-015-1291-0} {\bibfield  {journal}
  {\bibinfo  {journal} {J. Stat. Phys.}\ }\textbf {\bibinfo {volume} {160}},\
  \bibinfo {pages} {885} (\bibinfo {year} {2015})}\BibitemShut {NoStop}%
\bibitem [{\citenamefont {Krapivsky}\ \emph {et~al.}(2014)\citenamefont
  {Krapivsky}, \citenamefont {Mallick},\ and\ \citenamefont
  {Sadhu}}]{Krapivsky_PRL2014}%
  \BibitemOpen
  \bibfield  {author} {\bibinfo {author} {\bibfnamefont {P.~L.}\ \bibnamefont
  {Krapivsky}}, \bibinfo {author} {\bibfnamefont {K.}~\bibnamefont {Mallick}},\
  and\ \bibinfo {author} {\bibfnamefont {T.}~\bibnamefont {Sadhu}},\ }\bibfield
   {title} {\bibinfo {title} {Large deviations in single-file diffusion},\
  }\href {https://doi.org/10.1103/PhysRevLett.113.078101} {\bibfield  {journal}
  {\bibinfo  {journal} {Phys. Rev. Lett.}\ }\textbf {\bibinfo {volume} {113}},\
  \bibinfo {pages} {078101} (\bibinfo {year} {2014})}\BibitemShut {NoStop}%
\bibitem [{\citenamefont {Derrida}\ and\ \citenamefont
  {Gerschenfeld}(2009)}]{Derrida_JSP2009}%
  \BibitemOpen
  \bibfield  {author} {\bibinfo {author} {\bibfnamefont {B.}~\bibnamefont
  {Derrida}}\ and\ \bibinfo {author} {\bibfnamefont {A.}~\bibnamefont
  {Gerschenfeld}},\ }\bibfield  {title} {\bibinfo {title} {Current fluctuations
  in one dimensional diffusive systems with a step initial density profile},\
  }\href {https://doi.org/10.1007/s10955-009-9830-1} {\bibfield  {journal}
  {\bibinfo  {journal} {J. Stat. Phys.}\ }\textbf {\bibinfo {volume} {137}},\
  \bibinfo {pages} {978} (\bibinfo {year} {2009})}\BibitemShut {NoStop}%
\bibitem [{\citenamefont {Sadhu}\ and\ \citenamefont
  {Derrida}(2016)}]{Derrida_JSM2009}%
  \BibitemOpen
  \bibfield  {author} {\bibinfo {author} {\bibfnamefont {T.}~\bibnamefont
  {Sadhu}}\ and\ \bibinfo {author} {\bibfnamefont {B.}~\bibnamefont
  {Derrida}},\ }\bibfield  {title} {\bibinfo {title} {Correlations of the
  density and of the current in non-equilibrium diffusive systems},\ }\href
  {https://doi.org/10.1088/1742-5468/2016/11/113202} {\bibfield  {journal}
  {\bibinfo  {journal} {J. Stat. Mech.}\ }\textbf {\bibinfo {volume} {2016}},\
  \bibinfo {pages} {113202} (\bibinfo {year} {2016})}\BibitemShut {NoStop}%
\bibitem [{\citenamefont {Krapivsky}\ and\ \citenamefont
  {Meerson}(2012)}]{Krapivsky_PRE2012}%
  \BibitemOpen
  \bibfield  {author} {\bibinfo {author} {\bibfnamefont {P.~L.}\ \bibnamefont
  {Krapivsky}}\ and\ \bibinfo {author} {\bibfnamefont {B.}~\bibnamefont
  {Meerson}},\ }\bibfield  {title} {\bibinfo {title} {Fluctuations of current
  in nonstationary diffusive lattice gases},\ }\href
  {https://doi.org/10.1103/PhysRevE.86.031106} {\bibfield  {journal} {\bibinfo
  {journal} {Phys. Rev. E}\ }\textbf {\bibinfo {volume} {86}},\ \bibinfo
  {pages} {031106} (\bibinfo {year} {2012})}\BibitemShut {NoStop}%
\bibitem [{\citenamefont {Das}\ \emph {et~al.}(2017)\citenamefont {Das},
  \citenamefont {Kundu},\ and\ \citenamefont {Pradhan}}]{Das_PRE2017}%
  \BibitemOpen
  \bibfield  {author} {\bibinfo {author} {\bibfnamefont {A.}~\bibnamefont
  {Das}}, \bibinfo {author} {\bibfnamefont {A.}~\bibnamefont {Kundu}},\ and\
  \bibinfo {author} {\bibfnamefont {P.}~\bibnamefont {Pradhan}},\ }\bibfield
  {title} {\bibinfo {title} {Einstein relation and hydrodynamics of
  nonequilibrium mass transport processes},\ }\href
  {https://doi.org/10.1103/PhysRevE.95.062128} {\bibfield  {journal} {\bibinfo
  {journal} {Phys. Rev. E}\ }\textbf {\bibinfo {volume} {95}},\ \bibinfo
  {pages} {062128} (\bibinfo {year} {2017})}\BibitemShut {NoStop}%
\bibitem [{\citenamefont {Chakraborty}\ \emph {et~al.}(2020)\citenamefont
  {Chakraborty}, \citenamefont {Chakraborti}, \citenamefont {Das},\ and\
  \citenamefont {Pradhan}}]{Chakraborty_PRE2020}%
  \BibitemOpen
  \bibfield  {author} {\bibinfo {author} {\bibfnamefont {T.}~\bibnamefont
  {Chakraborty}}, \bibinfo {author} {\bibfnamefont {S.}~\bibnamefont
  {Chakraborti}}, \bibinfo {author} {\bibfnamefont {A.}~\bibnamefont {Das}},\
  and\ \bibinfo {author} {\bibfnamefont {P.}~\bibnamefont {Pradhan}},\
  }\bibfield  {title} {\bibinfo {title} {Hydrodynamics, superfluidity, and
  giant number fluctuations in a model of self-propelled particles},\ }\href
  {https://doi.org/10.1103/PhysRevE.101.052611} {\bibfield  {journal} {\bibinfo
   {journal} {Phys. Rev. E}\ }\textbf {\bibinfo {volume} {101}},\ \bibinfo
  {pages} {052611} (\bibinfo {year} {2020})}\BibitemShut {NoStop}%
\bibitem [{\citenamefont {Agranov}\ \emph {et~al.}(2023)\citenamefont
  {Agranov}, \citenamefont {Ro}, \citenamefont {Kafri},\ and\ \citenamefont
  {Lecomte}}]{Agranov_SPP2023}%
  \BibitemOpen
  \bibfield  {author} {\bibinfo {author} {\bibfnamefont {T.}~\bibnamefont
  {Agranov}}, \bibinfo {author} {\bibfnamefont {S.}~\bibnamefont {Ro}},
  \bibinfo {author} {\bibfnamefont {Y.}~\bibnamefont {Kafri}},\ and\ \bibinfo
  {author} {\bibfnamefont {V.}~\bibnamefont {Lecomte}},\ }\bibfield  {title}
  {\bibinfo {title} {Macroscopic fluctuation theory and current fluctuations in
  active lattice gases},\ }\href
  {https://doi.org/10.21468/SciPostPhys.14.3.045} {\bibfield  {journal}
  {\bibinfo  {journal} {SciPost Phys.}\ }\textbf {\bibinfo {volume} {14}},\
  \bibinfo {pages} {045} (\bibinfo {year} {2023})}\BibitemShut {NoStop}%
\bibitem [{\citenamefont {Jose}\ \emph {et~al.}(2023)\citenamefont {Jose},
  \citenamefont {Dandekar},\ and\ \citenamefont {Ramola}}]{Dandekar_JSM2023}%
  \BibitemOpen
  \bibfield  {author} {\bibinfo {author} {\bibfnamefont {S.}~\bibnamefont
  {Jose}}, \bibinfo {author} {\bibfnamefont {R.}~\bibnamefont {Dandekar}},\
  and\ \bibinfo {author} {\bibfnamefont {K.}~\bibnamefont {Ramola}},\
  }\bibfield  {title} {\bibinfo {title} {Current fluctuations in an interacting
  active lattice gas},\ }\href {https://doi.org/10.1088/1742-5468/aceb53}
  {\bibfield  {journal} {\bibinfo  {journal} {J. Stat. Mech.}\ }\textbf
  {\bibinfo {volume} {2023}},\ \bibinfo {pages} {083208} (\bibinfo {year}
  {2023})}\BibitemShut {NoStop}%
\bibitem [{\citenamefont {Chakraborty}\ and\ \citenamefont
  {Pradhan}(2024)}]{Tanmoy2023}%
  \BibitemOpen
  \bibfield  {author} {\bibinfo {author} {\bibfnamefont {T.}~\bibnamefont
  {Chakraborty}}\ and\ \bibinfo {author} {\bibfnamefont {P.}~\bibnamefont
  {Pradhan}},\ }\bibfield  {title} {\bibinfo {title} {Time-dependent properties
  of run-and-tumble particles. ii. current fluctuations},\ }\href
  {https://doi.org/10.1103/PhysRevE.109.044135} {\bibfield  {journal} {\bibinfo
   {journal} {Phys. Rev. E}\ }\textbf {\bibinfo {volume} {109}},\ \bibinfo
  {pages} {044135} (\bibinfo {year} {2024})}\BibitemShut {NoStop}%
\bibitem [{\citenamefont {Godr\`eche}(2003)}]{Godreche2003}%
  \BibitemOpen
  \bibfield  {author} {\bibinfo {author} {\bibfnamefont {C.}~\bibnamefont
  {Godr\`eche}},\ }\bibfield  {title} {\bibinfo {title} {Dynamics of
  condensation in zero-range processes},\ }\href
  {https://doi.org/10.1088/0305-4470/36/23/303} {\bibfield  {journal} {\bibinfo
   {journal} {J. Phys. A: Math. Gen.}\ }\textbf {\bibinfo {volume} {36}},\
  \bibinfo {pages} {6313} (\bibinfo {year} {2003})}\BibitemShut {NoStop}%
\bibitem [{\citenamefont {Rizkallah}\ \emph {et~al.}(2023)\citenamefont
  {Rizkallah}, \citenamefont {Grabsch}, \citenamefont {Illien},\ and\
  \citenamefont {Bénichou}}]{Rizkallah_JSM2023}%
  \BibitemOpen
  \bibfield  {author} {\bibinfo {author} {\bibfnamefont {P.}~\bibnamefont
  {Rizkallah}}, \bibinfo {author} {\bibfnamefont {A.}~\bibnamefont {Grabsch}},
  \bibinfo {author} {\bibfnamefont {P.}~\bibnamefont {Illien}},\ and\ \bibinfo
  {author} {\bibfnamefont {O.}~\bibnamefont {Bénichou}},\ }\bibfield  {title}
  {\bibinfo {title} {Duality relations in single-file diffusion},\ }\href
  {https://doi.org/10.1088/1742-5468/aca8fb} {\bibfield  {journal} {\bibinfo
  {journal} {J. Stat. Mech.}\ }\textbf {\bibinfo {volume} {2023}},\ \bibinfo
  {pages} {013202} (\bibinfo {year} {2023})}\BibitemShut {NoStop}%
\end{thebibliography}%

\begin{widetext}

\section{Appendix}

\subsection{Derivation of hydrodynamics of 2D SSEP in EM}

Our two-dimensional model with hard core occupancy can be described with occupation index of $\{i,j \}$-th site $\eta_{i,j}=1$ if the site is occupied or $\eta_{i,j}=0$ otherwise. The equilibrium limit of this model is when cell-cell interaction rate $p=0$ \cite{Chakraborti-PRR2024}. In this limit each particle jumps to any of the four direction with equal rate $1/4$ (see Fig.~\ref{Schem1}) and the model is called simple symmetric exclusion model (SSEP). Below we describe the steps to derive hydrodynamics for this model.

We denote $\bar{\eta}=1-\eta$ and the following relations are always true: $\eta^2=\eta$, $\bar{\eta}^2=\bar{\eta}$, and $\eta \bar{\eta}=0$. The time evolution probability rates of $\eta_{i,j}$ in an infinitesimal time $dt$ is given by
\begin{eqnarray}
\eta_{i,j}(t+dt) =
\left\{
\begin{array}{ll}
\bar{\eta}_{i,j}(t)            & {\rm prob.}~ {\cal Q} dt, \\
\eta_{i,j}(t)            & {\rm prob.}~ 1-{\cal Q} dt,
\end{array}
\right.
\label{ssep-rates}
\end{eqnarray}
where the stochastic weight ${\cal Q}$ has two parts: contribution from particle loss ${\cal Q}_1$ and contribution from particle gain ${\cal Q}_2$. In presence of a small bias $\textbf{F}=F_x \hat{x} + F_y \hat{y}$ the jump weights will have an additional factor $\left[1 \pm F_{x,y} \ell /2 \right]+{\cal O}(\textbf{F}^2)$ up to leading order term in $\textbf{F}$. 
For example, particle loss from $\{i,j \}$-th site can happen when $\{i,j \}$-th site is occupied and the neighboring four sites are unoccupied and thus
\be
{\cal Q}_1=\frac{1}{4} \eta_{i,j} \left[ \bar{\eta}_{i+1,j} \left( 1+\frac{F_x \ell}{2} \right) + \bar{\eta}_{i-1,j} \left( 1-\frac{F_x \ell}{2} \right) + \bar{\eta}_{i,j+1} \left( 1+\frac{F_y \ell}{2} \right) + \bar{\eta}_{i,j-1} \left( 1-\frac{F_y \ell}{2} \right) \right].
\label{Q1}
\ee
Similarly, we can write
\be
{\cal Q}_2=\frac{1}{4} \bar{\eta}_{i,j} \left[ \eta_{i+1,j} \left( 1-\frac{F_x \ell}{2} \right) + \eta_{i-1,j} \left( 1+\frac{F_x \ell}{2} \right) + \eta_{i,j+1} \left( 1-\frac{F_y \ell}{2} \right) + \eta_{i,j-1} \left( 1+\frac{F_y \ell}{2} \right) \right].
\label{Q2}
\ee

Taking average over realisations and considering local density $\rho_{i,j(t)}=\langle \eta_{i,j}(t) \rangle$, from Eq.~\eqref{ssep-rates} we can write the dynamical equation for $\rho_{i,j}(t)$ as
\be
\frac{d\rho_{i,j}}{dt} = \frac{\langle \eta_{i,j}(t+dt) \rangle -\langle \eta_{i,j}(t) \rangle}{dt} = \langle \bar{\eta}_{i,j}(t) {\cal Q}\rangle - \langle \eta_{i,j}(t) {\cal Q}\rangle.
\label{app1}
\ee
We insert Eqs.~\eqref{Q1} and \eqref{Q2} in Eq.~\eqref{app1} and use the relation $\eta \bar{\eta}=0$ to obtain 
\bea \nn
\frac{d\rho_{i,j}(t)}{dt} &=& \langle {\cal Q}_2 \rangle - \langle {\cal Q}_1 \rangle , \\ \nn
&=& \frac{1}{4}[(\rho_{i+1,j}+\rho_{i-1,j}-2\rho_{i,j}) + (\rho_{i,j+1}+\rho_{i,j-1}-2\rho_{i,j})] + \frac{F_x \ell}{8} \left( 1-2\rho_{i,j} \right)\left( \rho_{i-1,j}-\rho_{i+1,j} \right)  \\
&& + \frac{F_y \ell}{8} \left( 1-2\rho_{i,j} \right)\left( \rho_{i,j-1}-\rho_{i,j+1} \right). 
\label{app2}
\eea
Now, we go to the diffusive scaling limit by rescaling space $i \rightarrow x = i/{L}$, $j \rightarrow y = j/{L}$, time $t \rightarrow \tau = t/{L}^2$ and the lattice spacing  $\ell=1 \rightarrow 1/{L}$ and expand the density function $\rho$ around space $x$ and $y$ for small ${\cal O}(1/L)$ as following:
\bea
\rho_{i\pm 1,j} \equiv \rho \left( x \pm \frac{1}{{L}},y, \tau \right) = \rho(x,y, \tau) \pm \frac{1}{{L}} \frac{\partial \rho(x,y,\tau)}{\partial x} + \frac{1}{2 {L}^2} \frac{\partial^2 \rho(x,y, \tau)}{\partial x^2} + {\cal O} \left( \frac{1}{{L}^3} \right), \\
\rho_{i,j\pm 1} \equiv \rho \left( x,y \pm \frac{1}{{L}}, \tau \right) = \rho(x,y, \tau) \pm \frac{1}{{L}} \frac{\partial \rho(x,y,\tau)}{\partial y} + \frac{1}{2 {L}^2} \frac{\partial^2 \rho(x,y, \tau)}{\partial y^2} + {\cal O} \left( \frac{1}{{L}^3} \right).
\eea
We substitute this in Eq.~\eqref{app2} and keep terms up to order $1/{L}^2$ to obtain the hydrodynamic equation for the biased system as
\bea \nn
\frac{\partial {\rho}(x,y,\tau)}{{L}^2 \partial \tau} &=& \frac{1}{4 L^2} \left[ \frac{\partial ^2 \rho}{\partial x^2} + \frac{\partial ^2 \rho}{\partial y^2} \right] - \frac{1}{4 L^2} \left[ (1-2\rho) \frac{\partial \rho}{\partial x} F_x +(1-2\rho) \frac{\partial \rho}{\partial y} F_y \right] +{\cal O} \left( \frac{1}{{L}^3} \right), \\ \nn
\Rightarrow \frac{\partial {\rho}(x,y,\tau)}{\partial \tau} &=& \frac{1}{4} \left[ \frac{\partial ^2 \rho}{\partial x^2} + \frac{\partial ^2 \rho}{\partial y^2} \right] - \frac{1}{4} \left[ \frac{\partial}{\partial x} \{ \rho(1-\rho) \}  F_x + \frac{\partial}{\partial y} \{ \rho(1-\rho) \} F_y \right], \\
&=& -\bigtriangledown . \left[ -D(\rho) \bigtriangledown \rho + \chi(\rho) \textbf{F} \right],
\eea
with two density dependent transport coefficients $D(\rho)=1/4$ and $\chi(\rho)=\rho(1-\rho)/4$.

\subsection{Derivation of hydrodynamics in 1D UMTM}

Thus the mass transfer rates in the unbounded model can be written as:
\begin{eqnarray} \nonumber
m_i(t+dt) = ~~~~~~~~~~~~~~~~~~~~~~~~~~~~~~~~~~~~~~~~~~~~~~~~~~~~~~~\\
\left\{
\begin{array}{ll}
m_i(t) - 1            & {\rm prob.}~ q a_i dt, \\
m_i(t) + 1            & {\rm prob.}~ q a_{i-1} dt/2, \\
m_i(t) + 1            & {\rm prob.}~ q a_{i+1} dt/2, \\
m_i(t) - 1            & {\rm prob.}~ p a_i v(m_i) dt/2, \\
m_i(t) + 1            & {\rm prob.}~ p a_{i-1} v(m_{i-1})  dt/4, \\
m_i(t) + 1            & {\rm prob.}~ p a_{i+1} v(m_{i+1}) dt/4, \\
m_i(t)                & {\rm prob.}~ 1-\Sigma dt.
\end{array}
\right.
\label{unbounded-unbiased-s}
\end{eqnarray}

We now apply a small bias $F$ in the right direction and thus the jump weights will be modified as $c^F_{i j} =c_{i j} \left[1+ F(j-i) \ell /2 \right]+{\cal O}(F^2)$ up to leading order term in $F$. Incorporating this into Eq.~\eqref{unbounded-unbiased} we reach the mass update equation for a biased system as:
\begin{eqnarray}
m_i(t+dt) =
\left\{
\begin{array}{ll}
m_i(t) - 1            & {\rm prob.}~ (qa_i+pa_iv(m_i)) (1+ F \ell /2) dt/2, \\
m_i(t) - 1            & {\rm prob.}~ (qa_i+pa_iv(m_i)) (1- F \ell /2) dt/2, \\
m_i(t) + 1            & {\rm prob.}~ (qa_{i-1}+pa_{i-1}v(m_{i-1})) (1+ F \ell /2) dt/2, \\
m_i(t) + 1            & {\rm prob.}~ (qa_{i+1}+pa_{i+1}v(m_{i+1})) (1- F \ell /2) dt/2, \\
m_i(t)                & {\rm prob.}~ 1-\Sigma_F dt.
\end{array}
\right.
\label{unbounded-biased-s}
\end{eqnarray}
Now we show the master equation structure for the local mass density $\tilde{\rho}_i(t)=\langle m_i(t) \rangle$ as:
\begin{eqnarray}
[\tilde{\rho}_i(t+dt)-\tilde{\rho}_i(t)]/dt = (\tilde{\rho}_i-1)G_i + (\tilde{\rho}_i+1)[G_{i-1}(1+ F \ell /2)+G_{i+1}(1- F \ell /2)]/2 - \tilde{\rho}_i \Sigma_F ,
\end{eqnarray}
with $G_i=\langle g_i \rangle$ and $g_i=qa_i+pa_iv(m_i)$. After simplification, this gives
\begin{equation}
\frac{\partial \tilde{\rho}_i(t)}{\partial t} = \frac{1}{2} \left( G_{i-1}+G_{i+1}-2G_i \right) - \frac{F \ell}{4} \left( G_{i+1}-G_{i-1} \right).
\label{gradient-s}
\end{equation}
Now, going to the diffusive scaling limit by rescaling space $i \rightarrow x = i/\tilde{L}$, time $t \rightarrow \tau = t/\tilde{L}^2$ and the lattice spacing  $\ell=1 \rightarrow 1/\tilde{L}$, we obtain the hydrodynamic equation in the biased system as:
\begin{eqnarray} 
\frac{\partial \tilde{\rho}(x,\tau)}{\tilde{L}^2 \partial \tau} = \frac{1}{2} \left[ G \left( x-\frac{1}{\tilde{L}}, \tau \right) + G \left( x + \frac{1}{\tilde{L}}, \tau \right) - 2G(x, \tau) \right] - \frac{1}{4} \left[ G \left( x+\frac{1}{\tilde{L}}, \tau \right) - G\left( x - \frac{1}{\tilde{L}}, \tau \right) \right] \frac{F}{\tilde{L}}.
\label{grad-exp1}
\end{eqnarray}
Next we expand the function $G$ around space $x$ for small ${\cal O}(1/\tilde{L})$ as following:
\bea
 G \left( x \pm \frac{1}{\tilde{L}}, \tau \right) = G(\rho(x, \tau)) \pm \frac{1}{\tilde{L}} \frac{\partial G(\rho(x,\tau))}{\partial x} + \frac{1}{2 \tilde{L}^2} \frac{\partial^2 G(\rho(x, \tau))}{\partial x^2} + {\cal O} \left( \frac{1}{\tilde{L}^3} \right).
\eea
After substituting this expansion into Eq.~\eqref{grad-exp1} and keeping terms up to order $1/\tilde{L}^2$ we obtain the hydrodynamic equation for the biased system as
\bea
\frac{\partial \tilde{\rho}(x,\tau)}{\tilde{L}^2 \partial \tau} = \frac{1}{2} \left[ \frac{1}{\tilde{L}^2} \frac{\partial ^2 G(\tilde{\rho})}{\partial x^2} \right] - \frac{1}{4} \left[ \frac{2}{\tilde{L}} \frac{\partial G(\tilde{\rho})}{\partial x} \right] \frac{F}{\tilde{L}} +{\cal O} \left( \frac{1}{\tilde{L}^3} \right).
\eea
Therefore,
\bea
\nn
\frac{\partial \tilde{\rho}(x,\tau)}{\partial \tau} &=& - \frac{\partial}{\partial x} \left(-\frac{1}{2}\frac{\partial G(\tilde{\rho})}{\partial \tilde{\rho}} \frac{\partial \tilde{\rho}}{\partial x} + \frac{1}{2}G(\tilde{\rho}) F  \right) \\
&=&  - \frac{\partial}{\partial x} \left( -D(\tilde{\rho}) \frac{\partial \tilde{\rho}}{\partial x} + \chi(\tilde{\rho})F \right),
\label{drift-diffusion-s}
\eea
with the two transport coefficients
\begin{equation}
D(\tilde{\rho})=\frac{1}{2}\frac{\partial G(\tilde{\rho})}{\partial \tilde{\rho}} ~~~~~\mbox{and}~~~~ \chi(\tilde{\rho})=\frac{1}{2}G(\tilde{\rho}).
\label{transport-unbounded-s}
\end{equation}

\end{widetext}

\end{document}